\documentclass[prodmode,acmtompecs]{acmsmall} % Aptara syntax

\usepackage{rotating}
\usepackage{graphicx}
\usepackage[cmex10]{amsmath}	% It is recommended by IEEEtran_HOWTO to use amsmath in this way
\usepackage{amssymb}	% this is used for \mathbb
\usepackage{multirow}

\newtheorem{assumption}{\textbf{Assumption}}
\DeclareMathOperator*{\expectation}{\mathrm{E}} %

\newcommand{\reqvec}{\mathbf{q}}
\newcommand{\reqscalar}{q}
\newcommand{\ribvec}{\boldsymbol{\alpha}}
\newcommand{\ribscalar}{\alpha}

\newcommand{\feasibilityRegion}{F}
\newcommand{\fullUserSet}{N}
\newcommand{\taskScheduler}{\mathcal{X}}

\newcommand{\myComments}[1]{}	% this is used to comment out my comments in the text

\newif\ifinfocom
\infocomfalse

\newif\iftompecs
\tompecstrue

\newif\iftompecsonly
\tompecsonlyfalse

\newif\iftompecsextended
\tompecsextendedtrue

\newif\ifdissertation
\dissertationfalse

\newif\ifhuawei
\huaweifalse

\newcommand{\infocomStart}{\ifinfocom \myComments{Infocom: }}
\newcommand{\tompecsStart}{\iftompecs \myComments{TOMPECS version: }}
\newcommand{\tompecsonlyStart}{\iftompecsonly \myComments{TOMPECS only version: }}
\newcommand{\tompecsextendedStart}{\iftompecsextended  \myComments{TOMPECS extended version: }}
\newcommand{\dissertationStart}{\ifdissertation  \myComments{Dissertation version: }}
\newcommand{\huaweiStart}{\ifhuawei  \myComments{Huawei version: }}

\newcommand{\commentEnd}{\myComments{End}}

\newcommand{\add}[1]{#1}

\begin{document}

\markboth{Y. Du and G. de Veciana}{Scheduling for Cloud-Based Computing Systems to Support Soft Real-Time Applications}

% Title portion
\title{Scheduling for Cloud-Based Computing Systems to Support Soft Real-Time Applications}
\author{YUHUAN DU and GUSTAVO DE VECIANA
\affil{The  University of Texas at Austin}
}

\begin{abstract}
Cloud-based computing infrastructure provides an efficient means to support real-time processing workloads, e.g., virtualized base station processing, and collaborative video conferencing. This paper addresses resource allocation for a computing system with multiple resources supporting heterogeneous soft real-time applications subject to Quality of Service (QoS) constraints on failures to meet processing deadlines. We develop a general outer bound on the feasible QoS region for non-clairvoyant resource allocation policies, and an inner bound for a natural class of policies based on dynamically prioritizing applications' tasks by favoring those with the largest (QoS) deficits. This provides an avenue to study the efficiency of two natural resource allocation policies: (1) priority-based greedy task scheduling for applications with variable workloads, and (2) priority-based task selection and optimal scheduling for applications with deterministic workloads. The near-optimality of these simple policies emerges when task processing deadlines are relatively large and/or when the number of compute resources is large. Analysis and simulations show substantial resource savings for such policies over reservation-based designs.
\end{abstract}

\keywords{Soft real-time applications, cloud-computing, non-clairvoyant resource allocation, feasibility region, largest deficit first, greedy task scheduling, task selection and optimal scheduling, efficiency ratio}

\begin{bottomstuff}
\add{
This research was supported by Huawei Technologies Co. Ltd.}

\add{A conference version of this paper has been accepted to INFOCOM 2016. 
}
\end{bottomstuff}

\maketitle

\section{Introduction}
\label{sec_introduction}
The shift towards delivering compute platforms/services via cloud-based infrastructure is well on its way. 
An increasing number of the applications/services migrating to the cloud involve 
real-time computation with processing deadlines and where failure to meet the deadlines degrades user's Quality of Service (QoS). Such infrastructure allows one to reap 
the significant benefits of cloud computing, e.g., reduced cost of sharing computing, hoteling and cooling resources, 
along with increased reliability and energy efficiency. In this paper, we focus on Soft Real-Time (SRT) applications 
which can tolerate occasional violations of processing deadlines but still need to meet 
QoS or Service Level Agreements (SLA). 

\infocomStart
An example of such a platform is the Cloud-based Radio Access Network (CRAN) \cite{CMW, BLS14A, DuD14A} being considered for next generation cellular deployments. Instead of co-locating dedicated compute resources next to base station antennas, they virtualize compute resources for baseband processing. To do so, the received uplink signals associated with wireless subframes are sampled and sent from antennas to the cloud for timely decoding and processing such that
downlink signals requiring timely channel measurements, acknowledgements, etc., 
can be sent back to antennas for transmission. This process must happen within 
several milliseconds as determined by the cellular system standards. 
In this setting shared compute resources may occasionally fail to complete subframe processing on time, 
but this must happen infrequently, i.e., QoS/SLA requirements must be met. 
In fact, different tasks may have different QoS/SLA requirements. For example, failures in subframe baseband processing should be very infrequent whereas failures for tasks associated with channel measurement/estimation might be acceptable once every few subframes \cite{PCO}. 
Other SRT applications including collaborative video conferencing, multimedia processing, real-time control, augmented reality platforms, have similar characteristics. 
\commentEnd\fi

\tompecsStart
An example of such a platform is the Cloud-based Radio Access Network (CRAN) \cite{CMW,BLS14A,DuD14A} being considered for next generation cellular deployments. Instead of co-locating dedicated compute resources next to base station antennas, they virtualize compute resources for baseband processing. To do so, the received uplink signals associated with wireless subframes are sampled and sent from antennas to the cloud for timely decoding and processing such that
downlink signals requiring timely channel measurements, acknowledgements, etc., 
can be sent back to antennas for transmission. This process must happen within 
several milliseconds as determined by the cellular system standards. 
In this setting shared compute resources may occasionally fail to complete subframe processing on time, 
but this must happen infrequently, i.e., QoS/SLA requirements must be met. 
In fact, different tasks may have different QoS/SLA requirements. For example, failures in subframe baseband processing should be very infrequent whereas failures for tasks associated with channel measurement/estimation might be acceptable once every few subframes \cite{PCO}. 
Other SRT applications including multi-party collaborative video conferencing, multimedia processing, real-time control systems, augmented reality platforms, etc., have similar characteristics. 
\commentEnd\fi

\dissertationStart
Another example of an SRT service would be services associated with distributed multi-party collaborative video conferencing or educational applications. 
In this setting multiple participants at varied locations send their video and content 
to a processing center where it is combined, tailored, transcoded and sent back to distributed attendees 
in possibly distributed locations with different resolutions or points of view, etc.
In interactive settings, one must ensure small end-to-end delays and thus 
tight processing delays.  
%User can often tolerate a few glitches, as long as the glitches occur rarely. 
Still in many cases, it is acceptable that some video frames not be delivered
on time without substantially impacting user perceived quality of experience. 
\commentEnd\fi

\dissertationStart
Google glass and other augmented reality platforms share similar characteristics to the above examples. 
In such applications, a stream of local observations including video/sound could be sent to computing centers for processing, e.g., 
face and activity recognition, and results returned for display. To ensure ``fluidity'' the turnaround for such 
processing must be quite tight yet such applications may tolerate occasional failures in 
meeting processing deadlines if they are handled properly. 
\commentEnd
\fi

The computing infrastructure, e.g. \cite{VPK15A}, to support such applications may involve a large number of heterogeneous servers, e.g., 
various generations of processors, which themselves have multiple cores, special purpose hardware, shared memories/caches, etc. 
In other words, a complex collection of resources must be orchestrated to efficiently  
meet applications' SRT requirements. In this paper we focus on a single computing system, e.g., managed server/center, 
shared by a set of users, corresponding to SRT applications, that periodically generate workloads. The traditional management approach is to allocate dedicated resources to users to meet their QoS requirements. However, given the typical uncertainty in users' workloads and ``interference'' across shared resources, doing so typically involves over-provisioning. 

Computing systems today are engineered so as to permit prioritization 
of one user over another, e.g., production vs. non-production tasks, which in turn translates to priority in accessing
shared compute resources and/or memory. 
In this paper we consider resource allocation policies which can {\em dynamically} prioritize users in each period. 
Such dynamic prioritization of users would typically
reduce the required resources vs. static allocations, and is further flexible to changes in users' workload characteristics or QoS requirements.

%Figure~{\ref{figure_framework1}} displays our high-level view of the framework we consider.
%%It achieves this by dynamically prioritize users based on the task execution outcomes 
%%and/or the abstraction of the processing infrastructure and tasks' random workloads. 
%\begin{figure}[htp]
%  \centering
%  \includegraphics[width=0.4\textwidth]{Figures/framework.pdf}
%  \caption{High-level model for a computing system that supports dynamic user prioritization. }
%  \label{figure_framework1}
%\end{figure}

Given a set of users and a computing system, here are some key questions of interest:
\begin{longitem}
\item What QoS requirements are feasible? 
\item Can we design simple efficient resource allocation policies meeting users' QoS requirements and characterize the performance of these policies? 
\item Compared with dedicated resource allocation, what kinds of reductions in resource requirements can one expect from enabling dynamic resource sharing? 
\end{longitem}

In the sequel we will address these basic questions and more, but we first turn to related work. 

{\bf \em Related Work. }
\label{subsection_related_work}
There is a substantial body of work on scheduling real-time tasks. 
Starting with \cite{LiL73}, the community has established theoretical frameworks to study the scheduling of real-time applications where tasks are subject to hard deadlines, see e.g., \cite{Liu00b,DaB11,CFH04,Leu89}. The results typically assume worst case execution times/workloads and are too conservative for SRT applications.  

Different models have been introduced for the QoS needs of Soft Real-Time (SRT) applications. 
The work in \cite{HaR95,BeB97,Ram99} proposes the notion of $(m, k)$-firm deadlines requiring at least $m$ out of any $k$ consecutive tasks complete by their deadlines. 
But many services do not need such tight requirements and \add{the analytical results typically require deterministic workloads}. 
The authors in \cite{LLN87,HoK13b} consider imprecise computation models where each task consists of a mandatory part, which needs to complete by the deadline, and an optional part which improves the computational results. This is a reasonable model for tasks like artificial intelligence computation since additional optional iterations improve the results. However, many real-time tasks do not contain optional part and some of these tasks can miss the deadlines up to some degree. 
The work in \cite{LiA09} aims to guarantee bounded maximum deadline tardiness for all users. 
However, these frameworks and QoS models are not suitable for applications like CRAN and video conferencing where it is useless to process a task after its deadline and it is better to simply drop the task if it misses the deadline. 

This paper focuses on an SRT QoS model where a bound on the fraction of tasks completed on time is the QoS requirement. 
Such a model was first introduced in \cite{AtB98} where the authors propose a static allocation approach to meet such a QoS requirements. We shall use this as an evaluation benchmark. 
More recently, the authors in \cite{HoK13b,HoK12} adopt this QoS model to study a wireless access point supporting users that periodically generate packets which need to be transmitted within that period, and propose simple ``optimal'' scheduling policies. However, their results are limited to the setting where only one user can transmit at a time and where packet transmissions can be viewed as tasks with geometrically distributed workloads. 
% In this paper we propose a general framework involving multiple processing resources and generally distributed workloads. The model and results in \cite{HoK12} are a special case of this work. 

In this paper we consider prioritization policies that use the idea underlying longest-queue-first policies, whose performance has been studied in \cite{DiW06,JLS07,KWJ13} but in different settings. Moreover, the scheduling problem we consider is more than just one of ordering users according to a policy such as largest-deficit-first. We also need to design the task scheduler to allocate resources to tasks across a computing system's cores. 

Work on stochastic scheduling, e.g., \cite{BCS74,LLK93,Pin12b,AGG10,BDW86,AlS03} considers how to schedule a set of tasks with random workloads on multiple cores and aims to find a single schedule to minimize some objective function. Most of this type of work does not consider task completion deadlines and focuses on minimizing the expected completion time of the last task or the average expected completion time of all tasks. Moreover, such work typically assumes exponential workloads in order to get analytical results.  

Additional related work include those studying the mixing of real-time and non real-time traffic, see e.g., \cite{ShS01,JaS11,PaD07}, 
and those studying user/job management, see e.g., 
\cite{AAB00,MTH11,DeK14}. 

\dissertationStart
There is a substantial body of work on the study of scheduling users/services with streams of real-time tasks. 
Starting with \cite{LiL73A} the community has established solid frameworks for hard real-time services where all tasks need to be completed by deadlines and any violation is considered as a system failure \cite{Liu00b} \cite{DaB11A}. The results typically assume worst case execution times/workloads which can be much larger than the average case and cause inefficient utilization of the computing systems. However, for many soft real-time applications such as in the CRAN or video conferencing context, it is acceptable to miss the deadlines up to a limited failure rate as long as some QoS requirements are guaranteed. Moreover, these services typically have uncertainties in the workloads. 

In the literature, there are different models for the QoS requirement of soft real-time services. The work in \cite{HaR95A}, \cite{BeB97A} proposes the notion of $(m, k)$-firm deadlines which requires that for each user at least $m$ out of any $k$ consecutive tasks must be completed by their deadlines. But many services do not need such tight requirements. The authors in \cite{LLN87A}, \cite{HoK13bA} consider imprecise computation model where each task consists of a mandatory part which needs to be completed by the deadline and an optional part which improves the computational results but can always be dropped if the compute resource is limited. This is a reasonable model for tasks like artificial intelligence computation since additional optional iterations improve the results. However, many real-time tasks do not contain optional part that can always be dropped and the tasks can miss the deadlines up to some degree. The work in \cite{LiA09A} aims to guarantee bounded deadline tardiness for all users. The tardiness of a task is defined as the delay from the task deadline to the actual task completion time, and the tardiness for a user is the maximum tardiness of any of its tasks. But in some services like video conferencing and CRAN, it is useless to process a task after its deadline and it is better to simply drop it if the task misses the deadline. 

In this paper we focus on a soft real-time service model where users periodically generate tasks of random workloads and propose time-averaged number of tasks completed on time per period as QoS requirements. Such a QoS requirement model is first introduced in \cite{AtB98A} and recently adopted by \cite{HoK13bA} \cite{HoK12A} to study packet transmission in a wireless context. The authors in \cite{AtB98A} propose to meet the QoS requirements by allocating each user a fixed amount of time in each period where the amount of time is computed based on the workload distributions and the QoS requirements. But such an approach is not efficient when the workloads have large uncertainties. In this paper we use this as a benchmark to evaluate the performance of our approaches. 

The authors in \cite{HoK13bA} \cite{HoK12A} consider a wireless access point supporting users which periodically generate packets (tasks) that need to be transmitted within that period over independent heterogeneous unreliable wireless channels. 
The access point schedules only one user at each discrete time unit and the scheduled user transmits the packet successfully with some probability. 
For this setting, the authors propose simple scheduling policies which are shown to be ``optimal''. However, their results are restricted to the simplified setting. 
Since only one user can transmit at a time, the access point can be viewed as a simple single core. Considering packet transmissions as tasks then the transmission time corresponds to the tasks' workloads and is geometrically distributed, which has the memoryless property. In this paper we propose a general framework involving multiple cores and where users may generate generally distributed workloads, and we characterize the performance of some simple approaches. We will see that the model and results in these prior work is a special case of our work. 

Some additional work including \cite{ShS01A} \cite{JaS11A} studies mixing real-time and non real-time traffic, but considers restricted wireless models. The authors in \cite{DaB11A} \cite{SAA04} provide a more comprehensive review for real-time scheduling. 

In our proposed approach we use the idea of longest-queue-first policy whose performance is studied in \cite{DiW06} \cite{KWJ13A} but in different context. The work in \cite{DiW06} proposes sufficient conditions for the optimality of longest-queue-first policy in a generalized switch model which is different from our soft real-time framework in terms of task completion deadlines and random execution process in each period depending on the schedule and the workloads. 
The authors in \cite{KWJ13A} study the performance of a largest-deficit-first policy in a wireless setting where a set of links with QoS requirements may interfere with each other and at each time the system picks a set of links that do not interfere. But the mutually excluded interference model and assumption of constant service rate may not fit beyond wireless context. Moreover, the scheduling problem we will consider is more than just one of ordering users according to some policies like largest-deficit-first. We also need to design the task scheduler to coordinate the scheduling of tasks across cores. 

The problem of stochastic scheduling \cite{Pin12b} considers to schedule a set of tasks with random workloads on multiple cores and aims to find a single schedule to minimize some objective function. Most of these work focuses on minimizing the expected completion time of the last task or the average expected completion time of all tasks. And these work typically assume exponential workloads to get clean results. In our framework, in each period we also have to schedule tasks with random workloads across cores, but we need to finish tasks before deadlines and moreover we look at the long-term time-averaged effect of changing schedules from period to period rather than the effect of a single schedule. Also we look at more general workloads. 
\commentEnd
\fi

{\bf \em Our Contributions. }
In this paper, we consider a computing system consisting of multiple resources and study the scheduling of SRT users' random workloads subject to QoS constraints on timely task completions. To our knowledge, we are the first to give a theoretical 
characterization of the feasibility region for this general SRT framework and to consider performance and near-optimality of simple efficient scheduling policies. 
The contributions of this paper are threefold. 

First, we propose a general framework for SRT user scheduling on multiple resources, albeit we assume the workloads are New Better than Used in Expectation (NBUE) type. 
\dissertationStart
The authors believe it is a novel contribution to look at NBUE random workloads that characterize many workload distributions of interest and lead to nice results. 
\commentEnd\fi
In this framework, we develop an outer bound for the set of feasible QoS requirements for all possible non-clairvoyant resource allocation policies. 

Second, we study resource allocation policies which prioritize users based on Largest ``Deficit'' First (LDF) in each period and schedule tasks accordingly. We develop a general inner bound for the feasibility region for this class of policies. 
This enables us to study the efficiency of two policies: (1) LDF-based greedy task scheduling for users with variable workloads, and (2) LDF-based task selection and optimal scheduling for users with deterministic workloads. 
These simple policies are near-optimal when the deadlines are relatively large, and/or the number of resources is large. 
%We then prove simple results regarding the efficiency ratios and show near-optimality of two policies that combine the LDF user prioritization with different priority based task schedulers, which either greedily schedule tasks with variable workloads or combine task-selection and optimal scheduling for tasks with deterministic/low-variable workloads. 
%Based on that, we prove results regarding the efficiency ratios of two LDF-based policies and show their near-optimality in various scenarios. Specifically, the LDF policy which greedily processes tasks by priorities works well when period length is large, and the LDF policy with task-selection and optimal scheduling works well when users generate deterministic (constant) workloads or workloads with low variability. 

Finally, we evaluate the performance of the proposed policies in terms of the required number of resources to fulfill a given set of users' QoS requirements. 
We exhibit substantial savings versus a traditional reservation-based approach in various system settings. 
% We show that the resource savings can be substantial, e.g., $60$-$70\%$ under stringent QoS requirements and $80$-$90\%$ under more relaxed QoS requirements. 
We also discuss generalizations of our results when the resources have different processing speeds. 

{\bf \em Paper Organization. }
The paper is organized as follows: Section 2 introduces our system model and Section 3 describes a reservation-based approach and a general outer bound for the feasibility region. Section 4 discusses two prioritization-based policies and studies their efficiency ratios. Simulation results are exhibited in Section 5. Section 6 discusses generalizations and Section 7 concludes the paper. Some of the proofs are provided in the Appendix.

\section{System Model}
\label{sec_system_model}
We first introduce our user, system and QoS models. 

\subsection{Soft Real-Time (SRT) User Model}
We consider a computing system shared by a set of users $\fullUserSet = \{1, 2, \cdots, n\}$. 
The system operates over discrete periods $t = 1, 2, \cdots$. We denote by $\delta$ the length of a period.
In each period each of the $n$ users generates exactly one task. 
These tasks are available for processing at the beginning of the period, and need to complete by the end of the period. 
Tasks not completed on time are dropped, i.e., cannot be processed in subsequent periods. 
Here we assume a task is the unit of scheduling, i.e., a task cannot be processed in parallel. 

\begin{figure}[htp]
  \centering
  \includegraphics[width=0.4\textwidth]{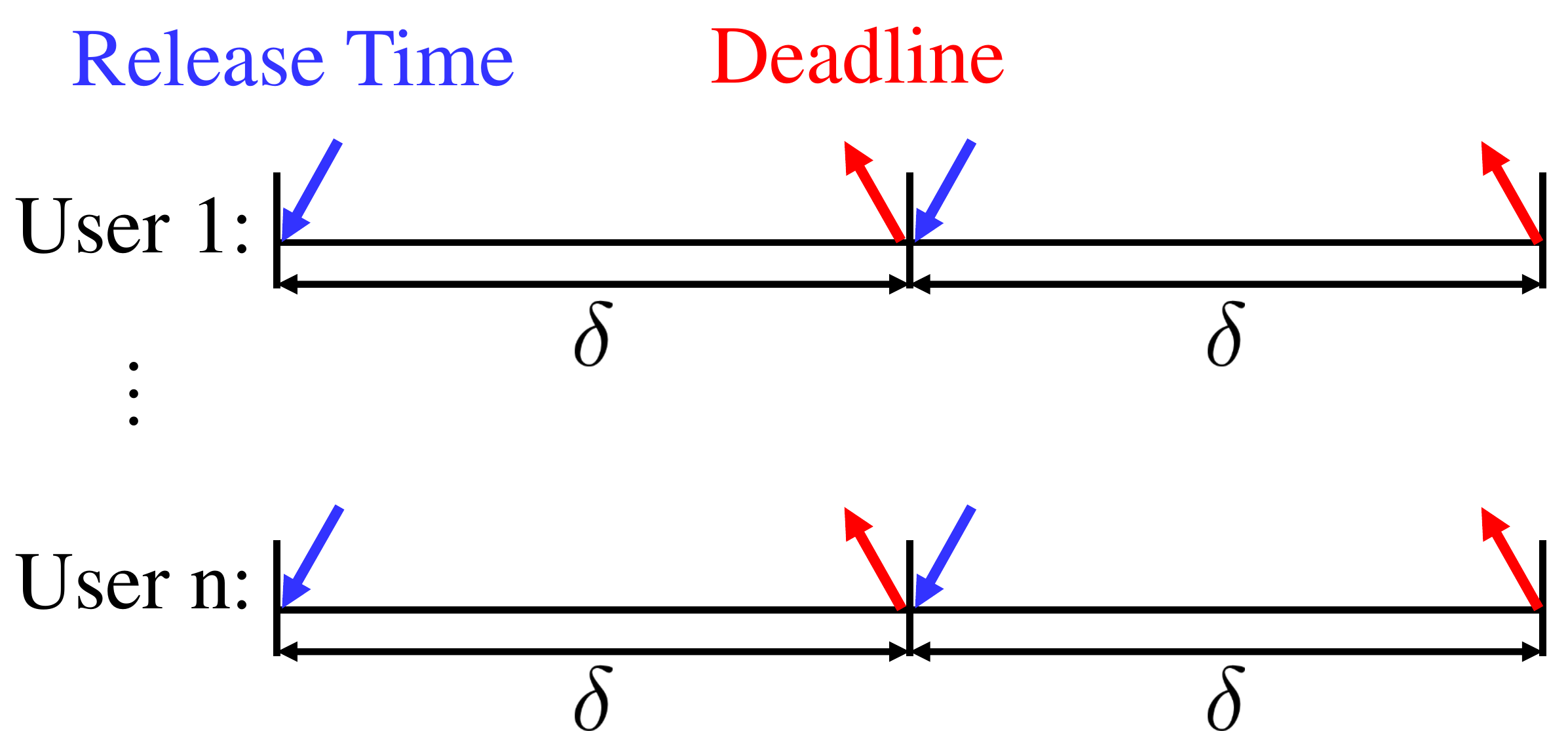}
  \caption{An example illustrating the task generation model. It shows the start and end points of periods. Each down and up arrow represents the release time and deadline of a task, respectively. }
  \label{figure_task_generation}
\end{figure}

\dissertationStart
We consider a system wherein $n$ users indexed from $1$ to $n$ share a centralized computing system. Let $\fullUserSet = \{1, 2, \cdots, n\}$ be the user set. 
The system operates in discrete time, over periods $t = 1, 2, \cdots$. We denote by $\delta$ the length of a period. 
The users generate streams of tasks periodically. Specifically in each period, each of the $n$ users generates exactly one task. 
These tasks are available for processing, i.e., released, at the beginning of the period, and need to complete by the end of the period. 
Tasks not completed on time are dropped, i.e., cannot be processed in subsequent periods. 
Figure~\ref{figure_task_generation} exhibits the above task generation process. 
Here we assume a task is the unit of scheduling, i.e., a task cannot be processed in parallel on multiple compute resources. 
In the sequel we will discuss generalizations where a task contains dependent sub-tasks. 
\commentEnd\fi

\dissertationStart
\commentEnd\fi

The workload of a task will refer to its resource requirement or service time. If a task's workload is large it may not be possible to complete on time. 
A task's workload is modeled by a random variable whose distribution captures variability in its resource requirement and/or uncertainty in the computing system, e.g., caused by memory contention across the cores. 
We assume task workloads for a given user are independent and identically distributed 
\tompecsStart
(i.i.d.)
\commentEnd\fi
across periods and workloads from different users are independent, possibly with different distributions. Let $W_i$ be a random variable denoting the workload of a task from user $i$ and let $\mu_i = \expectation[W_i]$. 
Next we introduce a further assumption on task workloads which seems reasonable for SRT users and will enable theoretical analysis. 

\begin{definition}
\infocomStart
A non-negative random variable $W$ is said to satisfy {\bf New Better than Used in Expectation (NBUE)} if for all $t > 0$, 
\commentEnd\fi
\tompecsStart
A non-negative random variable $W$ is said to satisfy {\bf New Better than Used in Expectation (NBUE)} if for all $t > 0$, 
\commentEnd\fi
\begin{align}
\label{align_NBUE}
\expectation[W - t | W > t] \leq \expectation[W]. 
\end{align}
\end{definition}
\dissertationStart
The exponential and geometric distributions are special NBUE distributions since they result in equality in the above for all $t>0$ and for all integers $t>0$, respectively. 
\commentEnd\fi
\dissertationStart
The notion of NBUE, see, e.g., \cite{MuS02b}\cite{ShS07b}, captures workload/lifetime distributions where the expected residual workload/lifetime of a task/device of age $t$ is no more than that of a new task/device. 
By \cite{ShS07b} the set of NBUE distributions is closed under convolution. Formally one can show the following results:
\begin{proposition}
\label{prop_NBUE_closure}
The NBUE property satisfies the following, 
\begin{itemize}
\item If two independent random variables $W, V$ are NBUE, then $W+V$ is NBUE. 
\item If a random variable $W$ is NBUE, then for all real numbers $c > 0$, $cW$ is NBUE. 
\end{itemize}
\end{proposition}
\myComments{end}
\fi
In this paper we shall assume all task workloads are NBUE. 

The NBUE property characterizes many workload distributions of interest. \cite{MuS02b} provides a discussion of NBUE distributions which include, but are not limited to, exponential, gamma with shape parameter $k \geq 1$ and deterministic distributions. 
A common class of distributions that are not NBUE is the heavy-tailed one. However since tasks need to complete within a period\footnote{In fact, we only require (\ref{align_NBUE}) to be true for $0 < t \leq \delta$. }, we are not likely to encounter tasks with such tails in the settings under consideration.  

\dissertationStart
The NBUE property characterizes many workload distributions of interest. By \cite{MuS02b} NBUE distributions include but are not limited to exponential distribution, gamma distribution with $k \geq 1$ and deterministic distribution (constant). 
Moreover, many distributions which are not NBUE can be closely approximated by NBUE distributions. For example, the normal distribution can be approximated by gamma distribution with large $k$. 
A common class of distributions that are not NBUE is the heavy-tailed one. However since in this setting tasks must complete within $\delta$, we only require (\ref{align_NBUE}) for $0 < t \leq \delta$ and thus are not likely to care about such tails in practice.  
\commentEnd\fi

We shall assume that each user $i$ has a QoS requirement given by a minimal long-term average number of tasks completed on time per period, denoted by $\reqscalar_i$ where $\reqscalar_i \in [0, 1]$. 
We let $\reqvec = (\reqscalar_1, \reqscalar_2, \cdots, \reqscalar_n)$ and 
assume $\reqscalar_i$'s are rational\footnote{All the results in this paper can be generalized to $\reqvec$'s with irrational values. For simplicity in the proof we do not consider that level of generality. }.

Let us consider some examples. An SRT user might correspond to the processing associated with a set of co-located cellular antennas in the CRAN context or an end user in video conferencing. 
Accordingly, the period $\delta$ would correspond to a wireless subframe or the length of a group of video frames, respectively. 
\dissertationStart
In CRAN each antenna generates a task associated with each subframe. Depending on the traffic and wireless channels, the tasks may take different CPU time and resources to process. 
\commentEnd\fi
For SRT users, it is generally useless to process a task after its deadline. For example, in video conferencing it is not desirable to display an out-of-date frame. This is why in this model tasks not completed on time are dropped. 
\infocomStart
In the extended version of this paper \cite{EXT}, 
\commentEnd\fi
\tompecsStart
In Section \ref{section_possible_generalizations}, 
\commentEnd\fi
we discuss possible generalizations where users may generate tasks with different periods and where a task may further consists of sub-tasks. 

\subsection{Computing Infrastructure}
A computing system can be very complex consisting of diverse, heterogeneous resources. 
In this paper, for simplicity of explanation we start with a computing system comprising of $m$ identical resources (cores)\textemdash a simple but relevant model. 
In Section 
\infocomStart
\ref{section_generalizations} 
\commentEnd\fi
\tompecsStart
\ref{section_possible_generalizations}
\commentEnd\fi
we discuss generalizations where cores have different processing speeds. 

Given $m$ identical cores, a task processed on any core requires the same processing time and each core can process only one task at a time. 
\dissertationStart
In this context, the workload of a task refers to the required core time to fully complete the task. 
\commentEnd\fi
In each period, the computing system dynamically schedules tasks according to a given strategy. Given the resource limit and the randomness of workloads, some tasks complete on time and some may fail. 

Unless otherwise specified we allow task preemption/migration, i.e., interrupting a task being processed and resuming later on the same/different core. We shall ignore the overheads of these operations. But in practice these operations involve context switching, and therefore, policies with minimal preemption and migration are desirable. 

A resource allocation policy is said to be {\em non-clairvoyant} if it does not make use of information regarding future events, such as tasks' workload realizations, which are not generally known until the tasks complete. However, a non-clairvoyant resource allocation policy may still have knowledge of a user's task workload distribution, which can be obtained from the history events or repeated experiments. We shall only consider non-clairvoyant resource allocation policies. 

In our model a ``core'' represents the minimum unit of compute resource such as physical computing core, specialized hardware, or hyper-thread as appropriate. The computing system could be a cloud-based cluster of machines or a centralized server with a collection of processors/cores. There are many possible non-clairvoyant resource allocation policies which may involve exploiting knowledge of workload distributions, exploiting history events, preempting tasks at appropriate times, dynamically prioritizing tasks, etc. 

\subsection{SRT QoS Feasibility}
Given a requirement vector $\reqvec$, a computing system and a non-clairvoyant resource allocation policy, how do we verify if $\reqvec$ is feasible?  
To keep track of the deficit among users' QoS requirements and actually completed tasks, for each user $i\in \fullUserSet$ and period $t+1$, we define\footnote{We truncate the deficit at $0$ via $[x]^+$ simply for the convenience of defining feasibility. Removing the truncation does not change the results in the paper. }
\begin{align}
\label{align_deficit}
X_i(t+1) = [X_i(t) + q_i - Y_i(t+1)]^+,
\end{align}
where $[x]^+ = \max[x, 0]$ and $Y_i(t+1)$ is an indicator random variable which takes value $1$ if user $i$'s task completes in period $t+1$. 
\infocomStart
The deficit vector $\mathbf{X}(t) = (X_1(t), X_2(t), \cdots, X_n(t))$ is a summary of the history of events up to period $t$. 
\commentEnd\fi
\tompecsStart
We let $\mathbf{X}(t) = (X_1(t), X_2(t), \cdots, X_n(t))$ denote the deficit vector. $\mathbf{X}(t)$ is a summary of the history of events up to period $t$. 
\commentEnd\fi

We shall say that the long-term QoS requirement $\reqscalar_i$ for user $i$ is met if and only if $X_i(t)$ is ``stable''. 
Formally, in this paper we consider non-clairvoyant resource allocation policies under which the process $\{\mathbf{X}(t)\}_{t\geq 1}$ is a Markov chain\footnote{
\infocomStart
All the results in this paper can be generalized to a broader range of non-clairvoyant policies and $\reqvec$'s with irrational values, see \cite{EXT}. 
\commentEnd\fi
\tompecsStart
All the results in this paper can be generalized to a broader range of non-clairvoyant resource allocation policies under which some variation of $\mathbf{X}(t)$ is a Markov chain. For example, if a resource allocation policy depends on the deficit vectors in the past two periods, then $\{(\mathbf{X}(t), \mathbf{X}(t+1))\}_{t\geq 1}$ is a Markov chain. For simplicity of explanation, we assume $\{ \mathbf{X}(t)\}_{t\geq 1}$ is a Markov chain.
\commentEnd\fi
}. 
We assume the initial state $\mathbf{X}(0)$, the QoS requirements $\reqvec$ and the policy make $\{\mathbf{X}(t)\}_{t\geq 1}$ an irreducible Markov chain. 

\begin{definition}
\label{defn_feasibility_pr}
We say the QoS requirement vector $\reqvec$ is {\bf feasible} if there exists a non-clairvoyant resource allocation policy $\eta$ under which the Markov chain $\{\mathbf{X}(t)\}_{t\geq 1}$ is positive recurrent, i.e., this policy fulfills $\reqvec$. 
We denote by $\feasibilityRegion_{\eta}$ the feasibility region of policy $\eta$, i.e., the set of QoS requirement vectors fulfilled by policy $\eta$. The union of $\feasibilityRegion_{\eta}$ over all allowable policies gives the {\bf system feasibility region} $\feasibilityRegion$. 
\end{definition}

We shall refer to this model as SRT-Multiple Identical Cores (SRT-MIC) with NBUE workloads and the aim is to devise non-clairvoyant resource allocation policies that fulfill $\mathbf{q}$.

\dissertationStart
Note that different users may require different QoS $q$. For example, in the CRAN context, soft real-time tasks include subframe baseband processing which requires $\reqscalar$ close to $1$, and channel measurements, which only need to be updated every few subframes and require low $\reqscalar$, probably $50\%$. 
\commentEnd\fi

In summary, the SRT-MIC model with NBUE workloads is an abstract system model which captures a family of systems supporting SRT users with random workloads. 
To summarize, the SRT-MIC model with NBUE workloads is parameterized by the number of cores $m$, number of users $n$, period length $\delta$, QoS requirements $\reqvec$, and the NBUE workload distributions.

\section{Reservation-Based Static Sharing and Outer Bound for the System Feasibility Region}
\dissertationStart
In the next few sections we will analyze and compare several resource allocation policies for the SRT-MIC systems with the aim of motivating good policies for more complicated practical systems.
\commentEnd\fi 
\add{Clearly simple policies like Earliest Deadline First (EDF) do not apply in our setting. Indeed in our problem statement all users generate tasks
which have the same deadline at the start of the scheduling interval. In fact in the sequel (see Section 6) we will see that even if users generate tasks with
different deadlines EDF performs poorly because it does not take the soft QoS requirements $\mathbf{q}$ into account.}

In this section we introduce a reservation-based policy and a general outer bound for the system feasibility region $\feasibilityRegion$ which applies to any non-clairvoyant resource allocation policy. These serve as benchmarks which enable us to evaluate the performance of the policies proposed in the sequel. 

%\begin{definition}
%A vector $\mathbf{x}$ is said to be {\bf dominated} by vector $\mathbf{y}$ if $x_i \leq y_i$ for all $i$ and is denoted by $\mathbf{x} \preceq \mathbf{y}$. We define $\mathbf{x} \prec \mathbf{y}$, $\mathbf{x} \succeq \mathbf{y}$ and $\mathbf{x} \succ \mathbf{y}$ in a similar manner. 
%\end{definition}

\subsection{Reservation-Based Static Sharing Policies}
\label{subsection_reservation_based_design}
A straightforward and commonly adopted approach to meet users' QoS requirements $\reqvec$ is to allocate dedicated resources, i.e., core time, to each user. For user $i$, with task workload $W_i$ and the requirement $q_i$, we let $w_i(q_i)$ represent the minimum core time reservation needed to ensure the requirement is met. Specifically, $w_i(q_i)$ is given by
$$
\Pr(W_i \leq w_i(q_i)) = q_i, 
$$
and thus, when $q_i$ is close to $1$, $w_i(q_i)$ will approach the worst-case workload for user $i$. 

\begin{figure}[htp]
  \centering
  \includegraphics[width=0.45\textwidth]{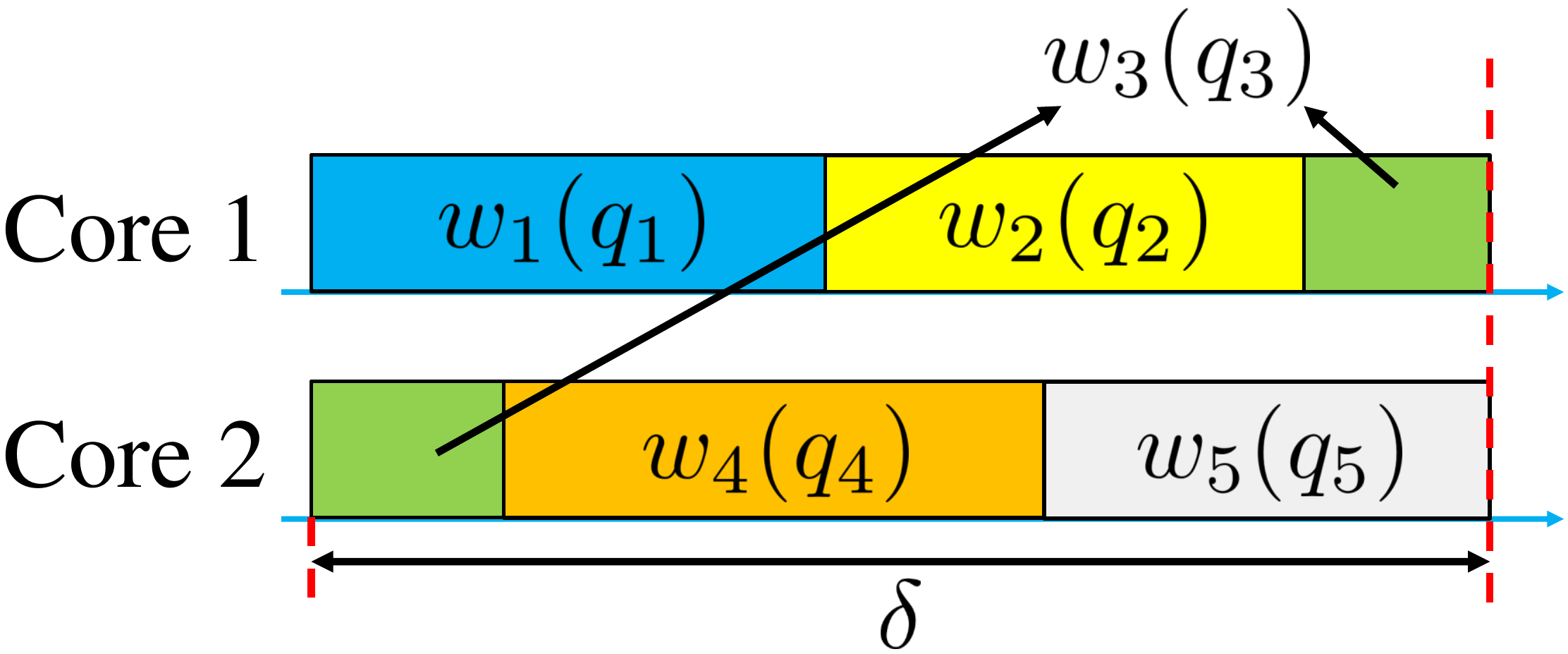}
  \caption{An example of the reservation-based approach. }
  \label{fig_reservation_based_design}
\end{figure}

{\em Reservation-based static sharing} policies allocate core time $w_i(q_i)$ to each user $i$ in each period and the tasks from users are only processed in the corresponding allocated time. Figure~{\ref{fig_reservation_based_design}} exhibits an example with 2 cores. Note that in this example User 3's task first executes on Core 2 and later continues on Core 1. Therefore, a reservation-based static sharing policy, although seemingly simple, can be aggressive in requiring task preemption/migration and knowledge of workload distributions to compute $w_i(\reqscalar_i)$ for all users. 

Note that since a task cannot be processed in parallel, if $w_i(q_i)$ exceeds the period length $\delta$, the requirement for user $i$ cannot be met. In this paper, we assume the task workloads and requirements $\reqvec$ are such that $w_i(q_i)$ is bounded by $\delta$. 

For a system with $m$ identical cores, the feasibility region $\feasibilityRegion_{\text{RB}}$ of reservation-based static sharing is given by
\begin{align}
\feasibilityRegion_{\text{RB}} = \{ \reqvec \in \mathbb R^n_+ ~|~ & \reqvec \preceq \mathbf{1}, \sum\limits_{i \in \fullUserSet} w_i(q_i) \leq m\delta, 	\notag \\
& w_i(q_i) \leq \delta, \forall i \in \fullUserSet \}, \label{align_F_RB_MIC}
\end{align}
where $\reqvec \preceq \mathbf{1}$ means $\reqscalar_i \leq 1$ for all $i \in \fullUserSet$. 
Clearly $\reqvec \preceq \mathbf{1}$ comes from the fact that each user generates only one task in each period. 

This approach was perhaps first proposed in \cite{AtB98A} and is also loosely used in reservation based schemes adopted in modern cloud infrastructure, see e.g., \cite{VPK15A}. 
Cores are not used efficiently under such a policy. When the realization of a task workload is smaller than the allocated time, the remaining time is wasted and cannot be used to process other real-time tasks. 
Typically , e.g. \cite{VPK15A}, the resources are then used to support best effort traffic. 

\subsection{Outer Bound for the System Feasibility Region $\feasibilityRegion$}
\label{subsection_outer_bound_for_F}
Ideally we aim to devise a policy that can fulfill all feasible QoS requirement vectors. More formally, a non-clairvoyant resource allocation policy $\eta$ is said to be {\em feasibility optimal} if its feasibility region $\feasibilityRegion_\eta$ is such that $\text{int}(\feasibilityRegion_\eta) \subseteq F \subseteq \text{cl}(\feasibilityRegion_\eta)$, where $\text{int}(\feasibilityRegion_\eta)$ and $\text{cl}(\feasibilityRegion_\eta)$ is the interior and closure of $\feasibilityRegion_\eta$, and thus is for practical purposes equivalent to the system feasibility region $\feasibilityRegion$. 

\dissertationStart
Ideally we wish to devise a policy that can fulfill all feasible QoS requirement vectors. More formally, a non-clairvoyant resource allocation policy $\eta$ is said to be {\em feasibility optimal} if its feasibility region $\feasibilityRegion_\eta$ is such that $\text{int}(\feasibilityRegion_\eta) \subseteq F \subseteq \text{cl}(\feasibilityRegion_\eta)$, where $\text{int}(\feasibilityRegion_\eta)$ and $\text{cl}(\feasibilityRegion_\eta)$ is the interior and closure of $\feasibilityRegion_\eta$. In other words, a policy $\eta$ is feasibility optimal if $\feasibilityRegion_\eta$ is different from the feasibility region $\feasibilityRegion$ by at most a boundary, and therefore, is equivalent to $\feasibilityRegion$ for practical purposes. 
\commentEnd\fi

\dissertationStart
Intuitively, a feasibility optimal non-clairvoyant design will orchestrate task scheduling across the $m$ cores. For example, the system might choose to process tasks with the maximum conditional probability of success to maximize the expected number of task completions in some periods while guaranteeing fairness and thus the required $\reqvec$ of users in other periods. 
\commentEnd\fi

Given the heterogeneity and randomness of tasks' workloads and the large number of possible non-clairvoyant resource allocation policies, a feasibility optimal policy is unknown except for very specific resource and workload models, see e.g., \cite{HoK12A}. To solve this and to provide a benchmark to evaluate other resource allocation policies, we develop a simple outer bound $R_{\text{OB}}$ for the system feasibility region $F$. 
Formally, we have the following theorem. 
\begin{theorem}
\label{thm_optimal_benchmark}
For the SRT-MIC  model with NBUE workloads, the system feasibility region $F$ is such that 
$$
F \subseteq R_{\textnormal{OB}} \equiv \{ \reqvec \in \mathbb R^n_+ ~|~ \reqvec \preceq \mathbf{1}, \sum\limits_{i\in \fullUserSet} q_i\mu_i \leq m\delta \}. 
$$
\end{theorem}

Intuitively, if $q_i$ tasks of user $i$ are completed each period, the expected time spent on user $i$ is roughly given by $q_i\mu_i$. To make $\reqvec$ feasible, the total time spent on all users $\sum\limits_{i\in \fullUserSet} q_i \mu_i$ cannot exceed the total available core time given by $m\delta$. 
This informal argument is perhaps deceptive. 
Note that in fact the expected time to complete the $q_i$ tasks for user $i$ in each period might be smaller than $q_i\mu_i$ since completed tasks might tend to have smaller workloads. This seems to imply that $m\delta$ could be smaller than $\sum\limits_{i\in \fullUserSet}\reqscalar_i \mu_i$ for some feasible $\reqvec$. This is where the NBUE assumption on workloads is critical to the result. 
\infocomStart
See Appendix \ref{appendix_pf_R_OB} for a detailed proof. 

This simple outer bound applies to any non-clairvoyant resource allocation policy in any specific SRT-MIC system with NBUE workload distributions. 
Note however the result does not necessarily hold for non-NBUE workloads. See the extended version of this paper \cite{EXT} for an illustrative example. 
\commentEnd\fi

\tompecsStart
Note this simple outer bound applies only to non-clairvoyant resource allocation policies for a specific SRT-MIC system with NBUE workload distributions. 
A formal proof of the theorem is given below. 

\begin{proof}
Given a feasible QoS requirement vector $\reqvec \preceq \mathbf{1}$, the goal is to show $\sum\limits_{i\in \fullUserSet} q_i\mu_i \leq m\delta$. 

Suppose $\reqvec$ is fulfilled by a non-clairvoyant resource allocation policy $\eta$, by definition $\{ \mathbf{X}(t)\}_{t \geq 1}$ is positive recurrent and therefore, there exists a stationary distribution. 
We consider a typical period where the deficit vector $\mathbf{X}(t)$ follows the stationary distribution and introduce further notation associated with period $t+1$. To simplify notation, we will suppress the period index in this proof. 

For each user $i$, we define $Y_i$ to be the indicator random variable that the task from user $i$ completes in a typical period. By the Ergodic Theorem, $\expectation[Y_i]$ also represents the time-averaged number of task completions per period for user $i$. If we view $X_i(t)$ as a queue, the average arrival $q_i$ should not exceed the average departure $\expectation[Y_i]$. For each user subset $S\subseteq \fullUserSet$, we define $U_S$ to be a random variable denoting the total core time spent on users in $S$ in a typical period. Clearly, $\expectation[U_S]$ cannot exceed the total available core time $m\delta$. To show $\sum\limits_{i\in \fullUserSet} q_i\mu_i \leq m\delta$, it suffices to show that $\sum\limits_{i\in \fullUserSet} \expectation[Y_i]\mu_i \leq \expectation[U_{\fullUserSet}]$. To that end we first develop an equation connecting $\sum\limits_{i\in \fullUserSet} \expectation[Y_i]\mu_i$ and $\expectation[U_{\fullUserSet}]$, and then use the NBUE assumption to show the inequality. 

We say a task is {\em unfinished} if it starts processing but does not complete in a given period. Let $A_i$ be the indicator random variable that user $i$'s task is unfinished in a typical period. Now if $Y_i + A_i = 1$ it indicates that user $i$'s task starts processing in the period though it may not have completed. 
For each user $i$, we further define $E_i = A_i(W_i - U_{\{i\}})$. Intuitively, $E_i$ represents the ``residual workloads for user $i$'s unfinished tasks''. 
Note that these random variables and their means depend on the policy $\eta$. 

For each user subset $S \subseteq \fullUserSet$, the total time spent on users in $S$ can be written as
$$
U_S = \sum\limits_{i\in S} (Y_i + A_i) W_i - \sum\limits_{i\in S} E_i, 
$$
and by taking expectations, we get
\begin{align}
\label{align_internal_exp}
\expectation[U_S] = \sum\limits_{i\in S} \expectation[(Y_i + A_i) W_i] - \sum\limits_{i\in S} \expectation[E_i]. 
\end{align}

Clearly $Y_i + A_i$, which indicates that user $i$'s task starts processing, is independent of $W_i$. Indeed this follows from the requirement that the resource allocation policy be non-clairvoyant, and the independence among users' task workloads. In a typical period under policy $\eta$, the event that user $i$'s task starts may depend on the workloads of others' tasks, but not on $W_i$. 

Note that although $Y_i + A_i$ is independent of $W_i$, in general $Y_i$ which indicates user $i$'s task completes may depend on $W_i$, i.e., $\expectation[Y_i W_i] \neq \expectation[Y_i]\mu_i$. To better understand this, consider an extreme example. If $W_i > \delta$, clearly the user $i$'s task cannot complete implying that $Y_i = 0$. Thus, $\expectation[Y_i | W_i > \delta] = 0 \neq \expectation[Y_i]$. 
Similarly, we can argue $A_i$ is not independent of $W_i$. 

Still given the independence of $Y_i + A_i$ and $W_i$, we have that
\begin{align*}
\expectation[(Y_i + A_i) W_i] & = \expectation[Y_i + A_i] \cdot \expectation[W_i] = (\expectation[Y_i] + \expectation[A_i])\mu_i. 
\end{align*}

So (\ref{align_internal_exp}) becomes
\begin{align}
\label{align_subset_time_equation}
\expectation[U_S] = \sum\limits_{i\in S}\expectation[Y_i]\mu_i + \sum\limits_{i\in S}\expectation[A_i] \mu_i - \sum\limits_{i\in S} \expectation[E_i]. 
\end{align}
This equation holds for all non-clairvoyant resource allocation policies and for all subsets of users $S\subseteq \fullUserSet$. 
%In the sequel we refer to this Subset Time Equation (STE).  

Now let $S = \fullUserSet$. To show $\sum\limits_{i\in \fullUserSet} \expectation[Y_i]\mu_i \leq \expectation[U_{\fullUserSet}]$, by (\ref{align_subset_time_equation}) it suffices to show $\expectation[A_i] \mu_i \geq \expectation[E_i]$ for all users $i\in \fullUserSet$. We will show this is true under the NBUE workload assumption in the discrete-time scenario and it is straightforward to generalize the proof to the continuous-time scenario. 

Suppose each period contains $\delta$ discrete time units. 
For all $i$ and for $c = 1, 2, \cdots, \delta$, we let $A_{i, c}$ denote the indicator random variable that user $i$'s task is unfinished and is processed for $c$ time units in a typical period. Clearly, $A_i = \sum\limits_{c = 1}^{\delta} A_{i, c}$ and $\expectation[A_{i, c}] = \Pr(A_{i, c} = 1)$. By the law of total probability, the expected residual workload $\expectation[E_i]$ for user $i$ can be written as
\begin{align}
\label{align_E_i_total_probability}
\expectation[E_i] = \sum\limits_{c = 1}^{\delta} \expectation[E_i | A_{i, c} = 1] \Pr(A_{i,c} = 1) = \sum\limits_{c = 1}^{\delta} \mu_{i,c} \expectation[A_{i, c}],
\end{align}
where $\mu_{i,c} = \expectation[W_i - c | W_i > c]$. 
This is because under the non-clairvoyant design the event $A_{i,c} = 1$ tells nothing about $W_i$ except that $W_i > c$. 

By the NBUE workload assumption we know that $\mu_{i, c} \leq \mu_i$ for $c > 0$ and therefore, we get the following inequality,
\begin{align}
\label{align_ret_smaller_than_mean}
\expectation[E_i] \leq \sum\limits_{c=1}^{\delta} \mu_i \expectation[A_{i, c}] = \mu_i \expectation[A_i]. 
\end{align}
Note that the equality holds if all users' task workloads follow geometric distributions (or exponential distributions in continuous-time scenario), possibly with different parameters. 

To summarize, by (\ref{align_subset_time_equation}) and (\ref{align_ret_smaller_than_mean}) we know that given a feasible requirement vector $\reqvec$, for all user subsets $S\subseteq \fullUserSet$,
\begin{align}
\label{align_required_workload_smaller_than_spent_time}
\sum\limits_{i\in S} \reqscalar_i \mu_i \leq \sum\limits_{i\in S} \expectation[Y_i] \mu_i \leq \expectation[U_S] \leq m\delta, 
\end{align}
which by letting $S = \fullUserSet$ implies
$
\sum\limits_{i\in \fullUserSet} \reqscalar_i \mu_i \leq m\delta, 
$
and thus, 
$
F \subseteq R_{\text{OB}}. 
$
\end{proof}

A key part of this argument is the inequality (\ref{align_required_workload_smaller_than_spent_time}), stating that for a feasible $\reqvec$ the ``effective'' workload $\sum\limits_{i\in S} \reqscalar_i \mu_i$ for any user subset $S$ should not exceed the total time spent on users in $S$, which is bounded by $m\delta$. 
This holds under the NBUE workload assumption but may not be true if users have non-NBUE task workloads. For example, suppose all users generate tasks with non-NBUE workloads as follows, 
$$
W_i = \left\{ 
   \begin{array}{l l}
     1 & \quad \text{with probability $0.5$}	\\
     9 & \quad \text{with probability $0.5$}. 	\\
   \end{array} \right.
$$
Clearly, the mean workload is $\mu_i = 5$. 
Let us consider such a policy. In each period, the system processes each task for exactly $1$ time unit and stops if the task does not complete because given its workload distribution we know this task will require $8$ more time units to complete. Suppose $m$ and $\delta$ is such that $m\delta = n$ and therefore, the system can process each task for $1$ time unit per period. Under such a policy we know $\reqscalar_i = 0.5$ for all user $i$ and the total time spent per period is $U_{\fullUserSet} = n$. Therefore, 
\begin{align*}
\sum\limits_{i\in \fullUserSet} \reqscalar_i\mu_i = 2.5n > n = \expectation[U_{\fullUserSet}] = m\delta, 
\end{align*}
which is not consistent with (\ref{align_required_workload_smaller_than_spent_time}) and Theorem \ref{thm_optimal_benchmark}. 
Non-NBUE workloads are beyond the scope of this paper. Yet for real-time computing workloads we expect NBUE to be a good assumption. 
\commentEnd\fi

\section{Largest Deficit First (LDF) Based Policies}
\begin{figure}[htp]
  \centering
  \includegraphics[width=0.55\textwidth]{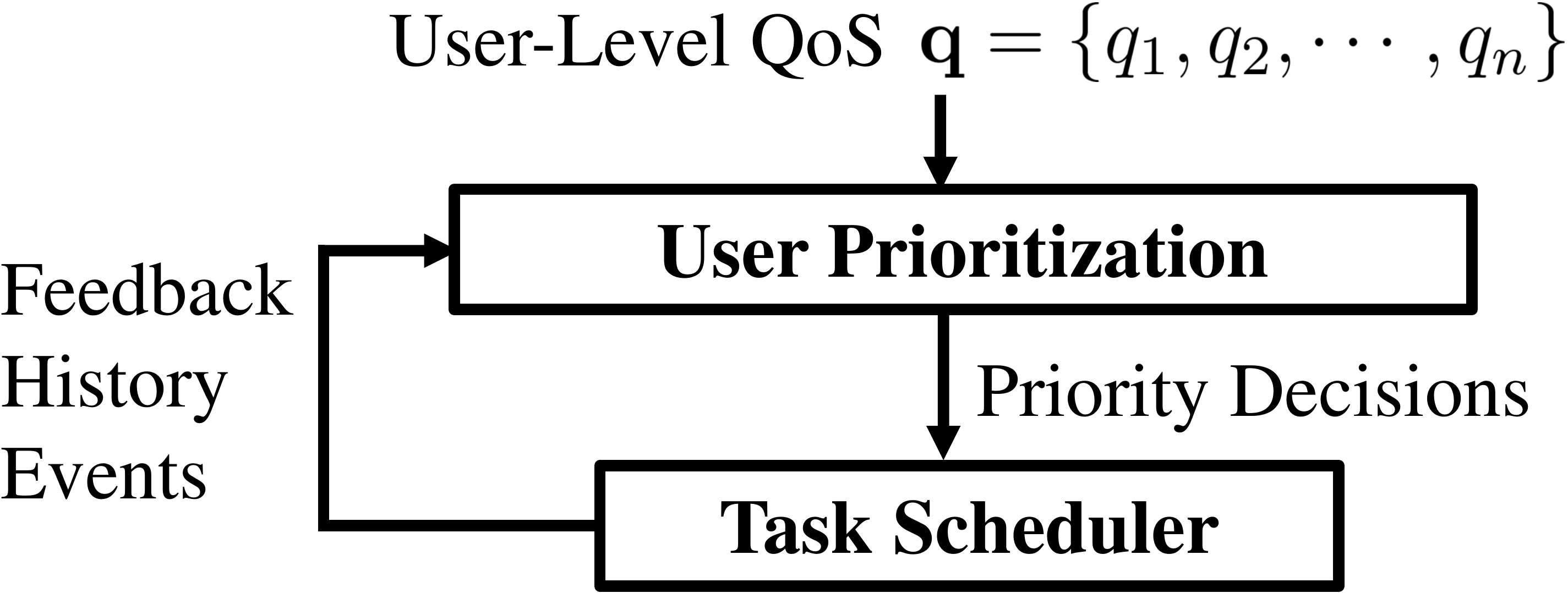}
  \caption{The framework for prioritization-based resource allocation policies. }
  \label{fig_prioritization_based_framework}
\end{figure}

Our aim is to devise a non-clairvoyant resource allocation policy that is easy to implement and whose feasibility region is near optimal. 
In this section we consider a specific class of policies, called {\em prioritization-based resource allocation} policies, which decompose resource allocation into two sub-problems, see Figure~{\ref{fig_prioritization_based_framework}}: 
\begin{enumerate}
\item \underline{User prioritization}: in each period the system dynamically prioritizes users based on the history of events. 
\item \underline{Task scheduler}: the system schedules users' tasks on cores based on their priorities. 
\end{enumerate}
There are still many options for each sub-problem. 
For example, task scheduling might be done greedily by simply scheduling the task with the highest priority, or using the priorities to first select a subset of tasks and then process that task subset via optimal scheduling policies. 

In this paper we shall prioritize users based on the Largest Deficit First (LDF) policy which is defined as follows. 

We let $\mathbf{d}=(d_1, d_2, \cdots, d_n)$ denote a {\em priority decision} where $d_k$ is the index of the user with $k^{\text{th}}$ highest priority and $D$ denote the set of all possible priority decisions.
\dissertationStart
and let $|D|$ be the number of possible decisions. We shall assume no restrictions on the set of allowable priority decisions, so $|D| = n!$
\commentEnd\fi

\begin{definition}
\label{defn_w_LDF}
The {\bf Largest Deficit First (LDF)} policy is such that, given the users' deficit vector $\mathbf{X}(t)$, the priority decision $\mathbf{d}$ for period $t+1$ is such that
$$
X_{d_1}(t) \geq X_{d_2}(t) \geq \cdots \geq X_{d_n}(t),
$$
with ties broken arbitrarily (possibly randomly). In other words, it sorts the deficits and assigns priorities accordingly. 
\end{definition}

The LDF user prioritization can be combined with different approaches of task scheduling. In the sequel we will explore such combinations and characterize their performance. 

\subsection{Inner Bound for Feasibility Region of LDF+$\taskScheduler$}
\label{subsection_LDF}
Given a task scheduling policy $\taskScheduler$, we let LDF+$\taskScheduler$ refer to the resource allocation policy that combines LDF user prioritization and task scheduler $\taskScheduler$.
In this subsection, we provide an inner bound for its feasibility region $\feasibilityRegion_{\text{LDF+}\taskScheduler}$.   

We first introduce some further notation. 
Given a task scheduler, in each period, the task completions depend on the selected priority decision. We let $p_i(\mathbf{d})$ denote the expected number of tasks completed in a period for user $i$ under priority decision $\mathbf{d}$ and let $\mathbf{p}(\mathbf{d}) = (p_1(\mathbf{d}), p_2(\mathbf{d}), \cdots, p_n(\mathbf{d}))$. Note that different task schedulers will correspond to different sets of vectors $P = \{\mathbf{p}(\mathbf{d}) | \mathbf{d} \in D\}$. 
We denote by $\mathbf{x} \succ \mathbf{0}$ a positive vector $\mathbf{x}$ with $x_i > 0$ for all $i\in \fullUserSet$. For all user subsets $S \subseteq \fullUserSet$, we let $|S|$ be the number of users in $S$ and we let $D(S)$ denote the set of all priority decisions that assign the highest $|S|$ priorities to users in $S$. 
The following theorem gives an inner bound on $\feasibilityRegion_{\text{LDF+$\taskScheduler$}}$. 
\begin{theorem}
\label{thm_R_IB}
Given a task scheduler $\taskScheduler$ and thus the $\taskScheduler$ dependent expected completion vectors $P = \{\mathbf{p}(\mathbf{d}) | \mathbf{d} \in D\}$, an inner bound for the feasibility region of the \tompecsStart resource allocation \commentEnd\fi policy LDF+$\taskScheduler$ is given by
$
\text{int}(R_{\textnormal{IB}}) \subseteq \feasibilityRegion_{\textnormal{LDF+$\taskScheduler$}}, 
$ 
where 
\begin{align*}
R_{\textnormal{IB}} \equiv \{\reqvec \in \mathbb R^n_{+} ~|~ & \exists \ribvec \succ \mathbf{0} \text{ such that } \forall S\subseteq \fullUserSet, \\
& \sum\limits_{i\in S} \ribscalar_i {\reqscalar}_i \leq \min\limits_{\mathbf{d} \in D(S)} \sum\limits_{i\in S} {\ribscalar}_i p_i(\mathbf{d})
\}. 
\end{align*}
\end{theorem}

Intuitively, $\reqvec$ is in $R_\text{IB}$ and is feasible under the LDF+$\taskScheduler$ policy if there is a weight vector $\ribvec \succ \mathbf{0}$ such that for any subset of users $S$, if the users in $S$ are given the highest priorities, the weighted sum of the requirements $\sum\limits_{i \in S}\ribscalar_i \reqscalar_i$ does not exceed the least weighted sum of the ``service rate'' $\sum\limits_{i\in S} {\ribscalar}_i p_i(\mathbf{d})$. Again, different task schedulers $\taskScheduler$ will have different vectors $P$ and thus different inner bounds $R_{\text{IB}}$. 
\infocomStart
For a detailed proof, see the extended version of this paper \cite{EXT}. 
\commentEnd\fi
\tompecsStart
A proof is provided in Appendix \ref{pf_theorem_R_IB}. 
Note that Theorem \ref{thm_R_IB} applies beyond the SRT-MIC model when the LDF policy is used but in a general setting where $\mathbf{p}(\mathbf{d})$ represent the expected payoffs under priority decision $\mathbf{d}$ and users require long-term time-averaged payoff $\reqvec$ per period. The LDF policy can also be generalized to a class of weighted LDF policies. This general result is further developed in \cite{DuD16E}. 
\commentEnd\fi

Next we explore specific task schedulers and use Theorem \ref{thm_R_IB} to study their performance. 

\dissertationStart
Note that Theorem \ref{thm_R_IB} applies beyond the SRT-MIC model when the LDF policy is used but in a general setting where $\mathbf{p}(\mathbf{d})$ represent the expected payoffs under priority decision $\mathbf{d}$ and users require long-term time-averaged payoff $\reqvec$ per period. The LDF policy can also be generalized to a class of weighted LDF policies. This general result is further developed in [?]. 
\commentEnd\fi

\subsection{Performance Analysis of LDF+Greedy Scheduling}
Given an LDF-based user priority decision in each period, a natural way to allocate resources is to greedily process tasks from highest to lowest priority. Specifically, to start by putting the $m$ tasks with the highest priority on the $m$ cores and, once one of these tasks completes, continue by processing the task with priority $m+1$ on the available core, etc.  

We let {\em LDF+Greedy} refer to the resource allocation policy that combines LDF and such a greedy task scheduler. Note this is easy to implement and does not require any a-priori knowledge of the tasks' workloads. Also this policy does not use task preemption or migration. 

Next we characterize the performance of LDF+Greedy. To that end, we introduce a metric called the efficiency ratio, see e.g., \cite{JLS07A}. The {\em efficiency ratio} of a non-clairvoyant resource allocation policy $\eta$ is defined as 
$$
\gamma_\eta = \sup\limits_{\gamma}\{\gamma | \gamma \feasibilityRegion \subseteq \feasibilityRegion_\eta \}. 
$$
Clearly $\gamma_\eta$ characterizes the performance gap between a policy $\eta$ and the best possible way of orchestrating the scheduling of multiple tasks across multiple cores. Also $\gamma_\eta$ equals to $1$ if and only if policy $\eta$ is feasibility optimal. 

\begin{theorem}
\label{thm_LDF_greedy_eff_ratio}
For the SRT-MIC model with NBUE workloads, the efficiency ratio of LDF+Greedy exceeds $\gamma_1$ where
$$
\gamma_1 = 1 - \frac{\max\limits_{i \in \fullUserSet} \mu_i}{\delta}. 
$$
\end{theorem}

% This result applies to any specific system that fits into the family of SRT-MIC  model with NBUE assumption. 
The intuition underlying this result is as follows. 
We say a task is {\em unfinished} if it starts processing but does not complete in a period.
The time spent on an unfinished task goes to waste since it does not contribute to a task completion. 
For LDF+Greedy, in one period, at most 1 task is unfinished per core and thus the wasted time on each core is expected to be less than $\max\limits_{i\in \fullUserSet} \mu_i$. Given the period is of length $\delta$, the gap between LDF+Greedy and optimality is bounded by $\frac{\max\limits_{i \in \fullUserSet} \mu_i}{\delta}$. 
Note that again this argument is deceptively simplified since unfinished tasks might tend to have larger workloads. 
Also as for Theorem \ref{thm_optimal_benchmark}, this result does not necessarily hold for non-NBUE workloads. 
\infocomStart
A sketch of the proof of this result using Theorem \ref{thm_R_IB} is included in Appendix \ref{appendix_pf_gamma_LDF_Greedy} and detailed proof is provided in the extended version of this paper \cite{EXT}. 
\commentEnd\fi
\tompecsStart
The formal proof is given below. 
\commentEnd\fi

\dissertationStart
We first provide an intuitive way to understand this theorem. 
We say a task is {\em unfinished} if it starts processing but does not complete in a period.
If a task is unfinished the time spent on the task goes to waste since it does not contribute to a task completion. 
For LDF+Greedy, in one period, since at most 1 task is unfinished per core, the wasted time on each core is expected to be less than $\max\limits_{i\in \fullUserSet} \mu_i$. Given the period length $\delta$, the gap between LDF+Greedy and optimality is bounded by $\frac{\max\limits_{i \in \fullUserSet} \mu_i}{\delta}$. 
Note that again this argument is deceptive since unfinished tasks might tend to have larger workloads. 
As for Theorem \ref{thm_optimal_benchmark}, this result does not necessarily hold for non-NBUE workloads. 
\commentEnd\fi

\tompecsStart
\begin{proof}
Given a requirement vector $\reqvec$ fulfilled by resource allocation policy $\eta$, by (\ref{align_required_workload_smaller_than_spent_time}) we know for all subsets of users $S \subseteq \fullUserSet$, 
$$
\sum\limits_{i\in S}\reqscalar_i \mu_i \leq \expectation[U_S], 
$$
where $\expectation[U_S]$ represents the time-averaged core time spent on users in $S$ per period under policy $\eta$. 

During each period, the total time $U_S$ spent on users in $S$ is bounded by the total task workload $\sum\limits_{i \in S} W_i$ of users in $S$ and the total available core time $m\delta$. 
We define $T_S = \min\left[\sum\limits_{i \in S} W_i, m\delta \right]$ and therefore, for all user subsets $S$, we have that
\begin{align}
\label{align_proof_LDF_greedy_necessary_condition}
\sum\limits_{i\in S}\reqscalar_i \mu_i \leq \expectation[U_S] \leq \expectation[T_S]. 
\end{align}

Thus, for a vector $\reqvec$ satisfying (\ref{align_proof_LDF_greedy_necessary_condition}) the aim to show $\gamma_\text{LDF+Greedy} \geq \gamma_1$ which is equivalent to showing $\gamma_1 \reqvec \in \text{cl}(\feasibilityRegion_\text{LDF+Greedy})$. By Theorem \ref{thm_R_IB}, it suffices to show that $\gamma_1 \reqvec \in R_{\text{IB}}$.
In LDF+Greedy, the expected vector $\mathbf{p}(\mathbf{d})$ described in Section \ref{subsection_LDF} represents the expected numbers of timely completions under greedy task scheduler under priority decision $\mathbf{d}$.  
Therefore, $\gamma_1 \reqvec \in R_{\text{IB}}$ follows if one can find a vector $\ribvec \succ \mathbf{0}$ such that for all $S \subseteq \fullUserSet$, 
$$
\sum\limits_{i\in S} \ribscalar_i \gamma_1 {\reqscalar}_i \leq \min\limits_{\mathbf{d} \in D(S)} \sum\limits_{i\in S} {\ribscalar}_i p_i(\mathbf{d}). 
$$
We will show $\ribvec = (\mu_1, \mu_2, \cdots, \mu_n) \succ \mathbf{0}$ satisfies the above condition. By (\ref{align_proof_LDF_greedy_necessary_condition}) it suffices to show for all $S$,
\begin{align*}
\gamma_1 \expectation[T_S] \leq \min\limits_{\mathbf{d} \in D(S)} \sum\limits_{i\in S} \mu_i p_i(\mathbf{d}), 
\end{align*}
which is equivalent to showing for any given user subset $S$ and priority decision $\mathbf{d} \in D(S)$ that
\begin{align}
\label{align_gamma_exp_T_S_leq}
\sum\limits_{i\in S} \mu_i p_i(\mathbf{d}) \geq \gamma_1 \expectation[T_S] = \expectation[T_S] - \frac{\max\limits_{i \in \fullUserSet} \mu_i}{\delta} \expectation[T_S].
\end{align}

First we rewrite $\sum\limits_{i\in S} \mu_i p_i(\mathbf{d})$ by similar approach used to obtain (\ref{align_subset_time_equation}). 
As in the proof of Theorem \ref{thm_optimal_benchmark}, for each subset of users $S\subseteq \fullUserSet$ and each user $i \in \fullUserSet$, 
we let $U_S(\mathbf{d})$, $A_i(\mathbf{d})$ and $E_i(\mathbf{d})$ denote the time spent on users in $S$, the indicator random variable that user $i$'s task is unfinished and the residual workload of user $i$'s unfinished tasks in a period under the greedy task scheduler with priority decision $\mathbf{d}$, respectively. 
% We let $a_i(\mathbf{d}) = \expectation[A_i(\mathbf{d})]$ and $e_i(\mathbf{d}) = \expectation[E_i(\mathbf{d})]$. 

By (\ref{align_subset_time_equation}), for the given $S$ and $\mathbf{d}$, we have that
\begin{align*}
% \label{align_gold_eqn_for_LDF_greedy}
\sum\limits_{i\in S}p_i(\mathbf{d})\mu_i = \expectation[U_S(\mathbf{d})] + \sum\limits_{i\in S} \expectation[E_i(\mathbf{d})] - \sum\limits_{i\in S}\expectation[A_i(\mathbf{d})] \mu_i .
\end{align*}

Now (\ref{align_gamma_exp_T_S_leq}) follows by showing that
\begin{align}
\label{align_exp_U_S_plus_E}
\expectation[U_S(\mathbf{d})] + \sum\limits_{i\in S} \expectation[E_i(\mathbf{d})] \geq \expectation[T_S]
\end{align}
and 
\begin{align}
\label{align_wasted_smaller_than_fraction_of_total}
\sum\limits_{i\in S}\expectation[A_i(\mathbf{d})] \mu_i \leq \frac{\max\limits_{i \in \fullUserSet} \mu_i}{\delta} \expectation[T_S], 
\end{align}
respectively. 

To demonstrate (\ref{align_exp_U_S_plus_E}), it suffices to show for each workload realization, 
$$
u_S(\mathbf{d}) + \sum\limits_{i\in S}e_i(\mathbf{d}) \geq t_S, 
$$
where $u_S(\mathbf{d}), e_i(\mathbf{d}), t_S$ are realizations of $U_S(\mathbf{d}), E_i(\mathbf{d}), T_S$, respectively. 

If $u_S(\mathbf{d}) = m\delta$, clearly $u_S(\mathbf{d}) + \sum\limits_{i\in S}e_i(\mathbf{d}) \geq m\delta \geq t_S$. Otherwise, $u_S(\mathbf{d}) < m\delta$. Since $\mathbf{d} \in D(S)$ assigns the highest priorities to users in $S$, by greedy task scheduler $u_S(\mathbf{d}) < m\delta$ implies that at the end of the period no task from users in $S$ is waiting to be scheduled, i.e., all tasks from users in $S$ start processing and therefore, $u_S(\mathbf{d}) + \sum\limits_{i\in S}e_i(\mathbf{d}) \geq \sum\limits_{i\in S}w_i \geq t_S$, where $w_i$ represents the realization of workload $W_i$. Therefore, (\ref{align_exp_U_S_plus_E}) is verified. 

Now it remains to show (\ref{align_wasted_smaller_than_fraction_of_total}). 
%Intuitively, $\sum\limits_{i\in S}\expectation[A_i(\mathbf{d})] \mu_i$ represents the wasted time spent on unfinished tasks and that should be bounded by $\frac{\max\limits_{i \in \fullUserSet} \mu_i}{\delta} \expectation[T_S]$. 
Clearly we have that
$$
\sum\limits_{i\in S}\expectation[A_i(\mathbf{d})] \mu_i \leq \max\limits_{i\in \fullUserSet} \mu_i \cdot \sum\limits_{i\in S} \expectation[A_i(\mathbf{d})]. 
$$
Thus, to demonstrate (\ref{align_wasted_smaller_than_fraction_of_total}) it suffices to show that
%\infocomStart
%$\sum\limits_{i\in S} \expectation[A_i(\mathbf{d})] \leq \frac{\expectation[T_S]}{\delta}$. 
%\commentEnd\fi
%\tompecsStart
\begin{align}
\label{align_num_unfinished_tasks_leq}
\sum\limits_{i\in S} \expectation[A_i(\mathbf{d})] \leq \frac{\expectation[T_S]}{\delta}. 
\end{align}
%\commentEnd\fi

We define $A_S(\mathbf{d}) = \sum\limits_{i\in S} A_i(\mathbf{d})$ to be the number of unfinished tasks in a period from users in $S$ under greedy task scheduler under priority decision $\mathbf{d}$. Since there are at most $m$ unfinished tasks, we have $A_S(\mathbf{d}) \leq m$. 

Under greedy task scheduling, for $\mathbf{d} \in D(S)$ we claim $A_S(\mathbf{d}) = k$ implies $T_S \geq k\delta$ for $k = 0, 1, \cdots, m$. This is true because $A_S(\mathbf{d}) = k$ means there are $k$ unfinished tasks on $k$ different cores, implying these $k$ cores are busy processing tasks from users in $S$ throughout the period.
Therefore, $\sum\limits_{i \in S}W_i \geq k\delta$ and thus $T_S \geq k\delta$. 
%Therefore, the total time $U_S(\mathbf{d})$ spent on users in $S$ exceeds $k\delta$. We have argued $T_S \geq U_S(\mathbf{d})$ and thus, $A_S(\mathbf{d}) = k$ implies $T_S \geq k\delta$. 

By this claim, we can get that
%\infocomStart
%\begin{align*}
%\expectation[T_S] & = \sum\limits_{k = 0}^{m} \expectation[T_S | A_S(\mathbf{d}) = k] \cdot \Pr(A_S(\mathbf{d}) = k)	\\
%& \geq \sum\limits_{k = 0}^{m} k\delta \cdot \Pr(A_S(\mathbf{d}) = k)	\\
%& = \delta \expectation[A_S(\mathbf{d})] = \delta \sum\limits_{i\in S} \expectation[A_i(\mathbf{d})]. 
%\end{align*}
%\commentEnd\fi
%\tompecsStart
\begin{align*}
\expectation[T_S] & = \sum\limits_{k = 0}^{m} \expectation[T_S | A_S(\mathbf{d}) = k] \cdot \Pr(A_S(\mathbf{d}) = k)	\\
& \geq \sum\limits_{k = 0}^{m} k\delta \cdot \Pr(A_S(\mathbf{d}) = k)	\\
& = \delta \sum\limits_{k = 0}^{m} k \cdot \Pr(A_S(\mathbf{d}) = k)	\\
& = \delta \expectation[A_S(\mathbf{d})]	\\
& = \delta \sum\limits_{i\in S} \expectation[A_i(\mathbf{d})]. 
\end{align*}
%\commentEnd\fi

This proves 
%\infocomStart
%(\ref{align_exp_U_S_plus_E}) 
%\commentEnd\fi
%\tompecsStart
(\ref{align_num_unfinished_tasks_leq}) 
%\commentEnd\fi
which in turn shows (\ref{align_gamma_exp_T_S_leq}) and therefore,
$
\gamma_1 \reqvec \in R_{\text{IB}} \subseteq \text{cl}(F_{\text{LDF+Greedy}}). 
$

\end{proof}
\commentEnd\fi

Theorem \ref{thm_LDF_greedy_eff_ratio} provides a lower bound on the efficiency ratio of LDF+Greedy, denoted by $\gamma_\text{LDF+Greedy}$. The bound is tight in the sense that for any $\epsilon > 0$, there exists an SRT-MIC  system with NBUE workloads such that $\gamma_{\text{LDF+Greedy}} < 1 - \frac{\max\limits_{i\in \fullUserSet} \mu_i}{\delta} + \epsilon$. 
\infocomStart
Such a system is also detailed in \cite{EXT}. 
\commentEnd\fi
\tompecsStart
Such a system is detailed in Appendix \ref{appendix_example_showing_theorem_LDF_Greedy_eff_ratio_is_tight}. 
\commentEnd\fi

\dissertationStart
Intuitively, $\frac{\max\limits_{i\in \fullUserSet} \mu_i}{\delta}$ represents the worst case ratio of wasted resources (core time) $m\max\limits_{i\in \fullUserSet} \mu_i$ under LDF+Greedy to the total available resources $m\delta$. 
\commentEnd\fi

It follows that if $\delta \gg \max\limits_{i\in \fullUserSet} \mu_i$, then $\gamma_1$ is close to $1$, i.e., LDF+Greedy is close to optimal. This is true when the task workloads are small relative to the core processing speed. 

However, when $\delta$ is comparable to $\max\limits_{i\in \fullUserSet} \mu_i$, the efficiency ratio lower bound $\gamma_1$ is small, 
although in some scenarios LDF+Greedy may still be efficient. For example, LDF+Greedy is feasibility optimal if the task workloads of all users follow the same exponential (or geometric) distribution, 
or prior work in \cite{HoK12}. 
This is due to the memoryless property of the exponential (or geometric) distribution. 
We omit the proof here. 
Still in some scenarios where we know more about the task workloads it is interesting to explore other simple policies that perform better than LDF+Greedy, especially when $\delta$ is comparable to the maximum mean workload. That motivates the discussion in the next subsection. 

% potential saving: reduce the title as before
\subsection{Performance Analysis of LDF+TS/LLREF Scheduling under Deterministic Workloads}
%\subsection{LDF+TS/LLREF under Deterministic Workloads}
\label{subsection_LDF_TS_LLREF}
In this subsection, we consider systems where users generate tasks with deterministic, but possibly different, workloads, i.e., $\Pr(W_i = \mu_i) = 1$ for all $i\in \fullUserSet$. 
For soft real-time users that can tolerate missing some deadlines, even if they generate tasks with deterministic workloads, one can still intentionally drop a fraction of tasks in each period while guaranteeing the users' long-term QoS requirements. 
Selecting a subset of tasks to be processed in each period is like a bin backing problem. And to fulfill the long-term soft QoS requirements, one need to dynamically change or rotate the selected task subset. 
%Even for deterministic workloads, the scheduling of tasks in different periods to meet soft QoS requirements is not straightforward. 

Note deterministic workloads satisfy the NBUE property. Also note that for deterministic workloads, non-clairvoyant policies have knowledge of workload realizations. 
We shall once again prioritize users using LDF prioritization. 
Intuitively, the greedy task scheduler wastes time on multiple cores if multiple tasks are unfinished at the end of a period, so we will devise a task scheduler that orchestrates across cores so as to ``reduce'' wasted core time to finish more tasks. 

For deterministic workloads, one can assess how many tasks one can complete prior to initiating processing. 
Indeed, it is intuitive, and established in \cite{CRJ06}, that one can complete all tasks in a user subset $S$ in a period by some optimal scheduling if and only if $\sum\limits_{i \in S} \mu_i \leq m\delta$. We consider one such optimal algorithm: Largest Local Remaining Execution time First (LLREF) \cite{CRJ06}. Let us briefly describe how LLREF\footnote{LLREF is defined to be applicable in more general settings where users might generate tasks with different period. We will discuss this in Section \ref{section_possible_generalizations}. } would work in the SRT-MIC model and then introduce a task scheduler that combines the idea of task selection and LLREF scheduling. 

To that end we introduce some terminology used in \cite{CRJ06}. Consider a period starting at time $t\delta$ and ending at time $(t+1)\delta$ , at any time $\tau \in [t\delta, (t+1)\delta]$, the {\em Local Remaining Execution time (LRE)} of user $i$ is defined as the remaining time needed to complete its task. The LRE decrements as the task is processed. 
Further, the {\em laxity} of user $i$ is defined as the remaining time before the deadline of user $i$'s task, i.e., $(t+1)\delta - \tau$, minus the current LRE of user $i$. Thus, if some user has zero laxity at some time, one needs to start processing the task immediately to complete it by its deadline. 

\begin{definition}
\label{defn_LLREF}
For the SRT-MIC model with deterministic workloads, the {\bf Largest Local Remaining Execution time First (LLREF)} policy is such that, given a selected user subset $S$ for the period, it does the following: 
\begin{enumerate}
\item At the beginning of the period, $m$ tasks associated with users in $S$ are chosen to be processed according to largest LRE first.
\item When a running task completes, or a non-running task reaches a state where it has zero laxity, again the $m$ tasks in $S$ with largest 
\infocomStart
LRE
\commentEnd\fi
\tompecsStart
local remaining execution time
\commentEnd\fi
are selected to be processed. 
\end{enumerate}
\end{definition}
Note that the LLREF policy uses task preemption and possibly migration. A review of variants of LLREF aimed at reducing task preemptions is provided in \cite{DaB11A}. 

\begin{figure}[htp]
  \centering
  \includegraphics[width=0.55\textwidth]{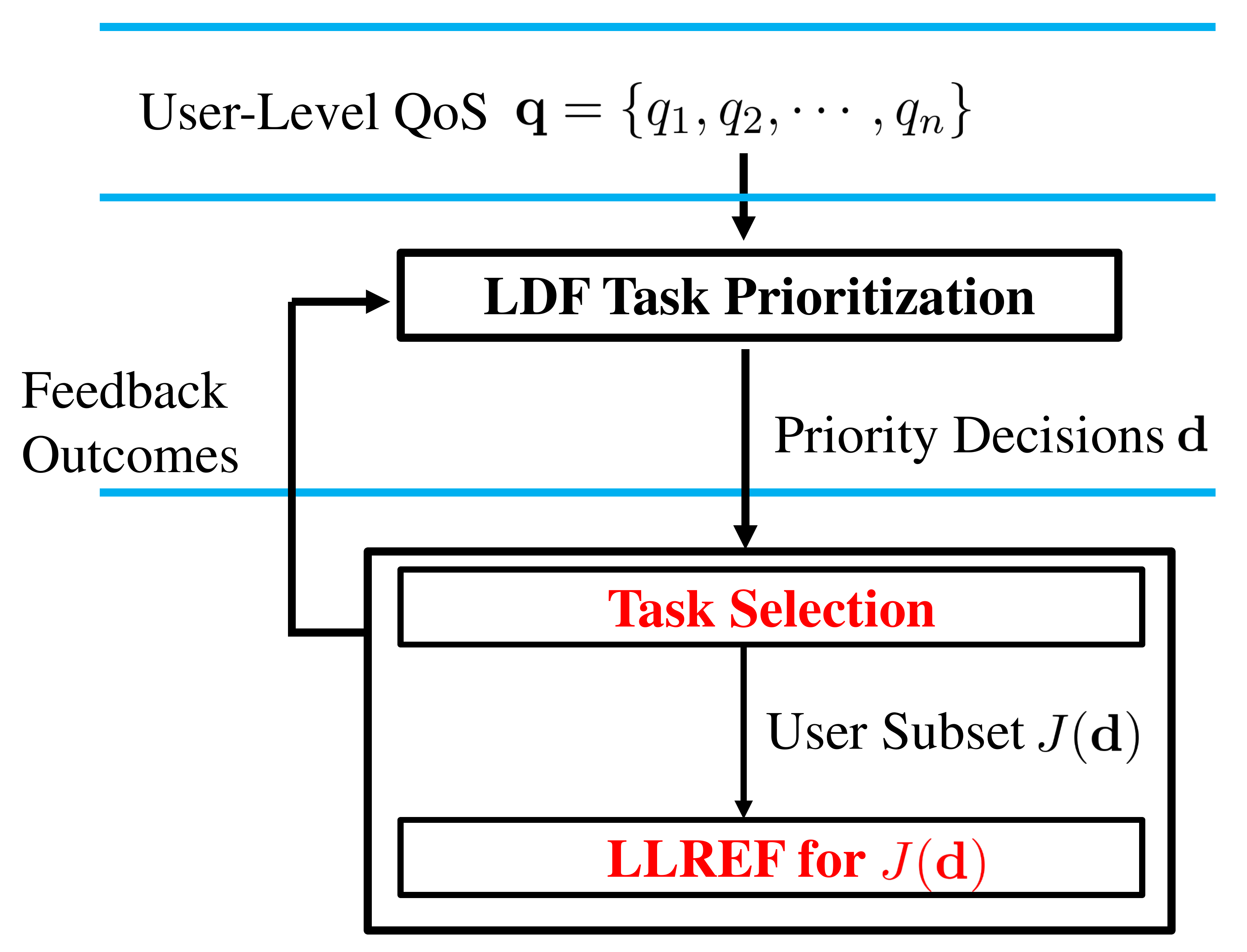}
  \caption{The framework for LDF+TS/LLREF design. }
  \label{fig_LDF_TS_LLREF_framework}
\end{figure}

\dissertationStart
Now we propose the following task scheduler. The framework is exhibited in Figure~{\ref{fig_LDF_TS_LLREF_framework}}. 
\commentEnd\fi

\begin{definition}
\label{defn_TS_LLREF}
The {\bf Task Selection/LLREF (TS/LLREF)} task scheduler is such that, given the user priority decision $\mathbf{d}$ for a period, it does the following: 
\begin{enumerate}
\item Task selection: it greedily selects users based on $\mathbf{d}$ until the sum workload exceeds $m\delta$. More formally, it selects
\begin{align}
\label{align_j_d}
j(\mathbf{d}) = \max\Big\{j | \sum\limits_{i = 1}^{j} \mu_{d_i} \leq m\delta \Big\}. 
\end{align}
Let $J(\mathbf{d}) = \{d_1, d_2, \cdots, d_{j(\mathbf{d})}\}$ represent the selected user subset. 
\dissertationStart
subset\footnote{One trivial way to extend the task selection step is to continue checking users greedily according to $\mathbf{d}$ and adding users to $J(\mathbf{d})$ while guaranteeing $\sum\limits_{i\in J(\mathbf{d})} \mu_i \leq m\delta$. But that does not improve the results in the sequel and therefore, we do not discuss this extension. }. 
\commentEnd\fi
\item LLREF for $J(\mathbf{d})$: the system uses LLREF scheduling for tasks in $J(\mathbf{d})$ in this period. 
\end{enumerate}
\end{definition}
By \cite{CRJ06A}, it follows that all tasks from $J(\mathbf{d})$ will complete. 

\dissertationStart
\commentEnd\fi

Paralleling Theorem \ref{thm_LDF_greedy_eff_ratio}, we have the following result for the LDF+TS/LLREF resource allocation, i.e., the combination of LDF user prioritization and TS/LLREF task scheduling. 
\dissertationStart
The framework of LDF+TS/LLREF is exhibited in Figure{~\ref{fig_LDF_TS_LLREF_framework}}. 
\commentEnd\fi

\begin{theorem}
\label{thm_LDF_TS_LLREF_eff_ratio}
For the SRT-MIC model with deterministic workloads, the efficiency ratio of LDF+TS/LLREF exceeds $\gamma_2$ where
$$
\gamma_2 = 1 - \frac{\max\limits_{i \in \fullUserSet} \mu_i}{m\delta}. 
$$
\end{theorem}

Intuitively, under TS/LLREF, the task selection rule guarantees that in any given period the wasted time $m\delta - \sum\limits_{i\in J(\mathbf{d})} \mu_i$ is less than $\max\limits_{i\in \fullUserSet} \mu_i$. Given the total available core time $m\delta$, the gap between LDF+TS/LLREF and optimality is again bounded by the fraction of wasted time, i.e.,  $\frac{\max\limits_{i \in \fullUserSet} \mu_i}{m\delta}$. A formal proof of this result is similar to that of Theorem \ref{thm_LDF_greedy_eff_ratio} and is provided 
\infocomStart
in the extended version of this paper \cite{EXT}. 
\commentEnd\fi
\tompecsStart
in Appendix \ref{appendix_pf_thm_LDF_TS_LLREF_eff_ratio}. 
\commentEnd\fi

The efficiency ratio lower bound $\gamma_2$ in this theorem is better than $\gamma_1$ obtained in Theorem \ref{thm_LDF_greedy_eff_ratio}, specifically the dependence on $m$ is much stronger. For a system with a large number of cores $m$, $\gamma_2$ is close to $1$, i.e., LDF+TS/LLREF is close to feasibility optimal even if $\delta$ is comparable to $\max\limits_{i\in \fullUserSet} \mu_i$. 

\dissertationStart
Under the more general NBUE (but not deterministic) workload assumption, if we somehow know the workload realizations at the beginning of each period, 
we can still adopt the LDF+TS/LLREF policy where the task selection is based on workload realizations. By similar proof as that for Theorem \ref{thm_LDF_TS_LLREF_eff_ratio} we can get the following corollary:
\begin{corollary}
For the SRT-MIC model with random NBUE workloads, if a QoS requirement vector $\reqvec$ can be fulfilled by some non-clairvoyant design, then under LDF+TS/LLREF which is clairvoyant $\gamma_2 \reqvec \in \text{cl}(F_{\text{LDF+TS/LLREF}})$, where
$$
\gamma_2 = 1 - \frac{\max\limits_{i \in \fullUserSet} \mu_i}{m\delta}. 
$$
\end{corollary}

In this corollary we are only comparing with non-clairvoyant designs, but note that in this scenario the LDF+TS/LLREF is a clairvoyant design because it requires knowledge of workload realizations, which can be hard in practice. 
\commentEnd\fi

Although LDF+TS/LLREF is designed for deterministic workloads, we envisage it will work well for workloads with small variability by using the expected workload, or some more sophisticated workload estimation $w_i^{\text{est}}$. Specifically, TS makes selections based on $w_i^{\text{est}}$ and LLREF computes local remaining execution time and laxity by assuming $W_i = w_i^{\text{est}}$. 
Note that this heuristic LDF+TS/LLREF is still non-clairvoyant. This will be explored in the simulation section. 

\dissertationStart
This heuristic LDF+TS/LLREF is still non-clairvoyant but picking appropriate $w_i^\text{est}$ is key to the performance. 
On one hand big $w_i^\text{est}$ reduces the number of tasks selected per period and on the other hand small $w_i^{\text{est}}$ may cause user $i$'s task fail to complete. 
In the simulation section, we will show such a policy performs well for NBUE workloads that are not deterministic but have small variability. 
\commentEnd\fi

\subsection{Resource Requirements}

So far we have analytically characterized the efficiency ratios of two LDF-based resource allocation policies. Another metric of interest is the resource requirements in terms of the number of cores $m$ needed to fulfill a set of users' QoS requirements. To that end in this subsection we shall explore the required $m$ given $n$, $\delta$, the random workload distributions and the requirement vector $\reqvec$. A policy that requires a smaller $m$ is better in that it saves compute resources and/or energy. 

\subsubsection{Resource Requirements for Reservation-Based Static Sharing}
~

Based on the definition of $F_{\text{RB}}$ in \ref{subsection_reservation_based_design}, the required number of cores to fulfill the users' QoS requirements $\reqvec$ under reservation-based static sharing is given by
\begin{align}
\label{align_m_RB}
m_{\text{RB}} = \Big\lceil \frac{\sum\limits_{i\in \fullUserSet}w_i(q_i)}{\delta} \Big\rceil, 
\end{align}
where $\lceil x \rceil$ is the ceiling of $x$. 

\subsubsection{Lower Bound on Resource Requirements}
~

For any non-clairvoyant resource allocation policy $\eta$, we let $m_{\eta}$ denote the required number of cores to fulfill users' QoS requirements under policy $\eta$. 
By Theorem \ref{thm_optimal_benchmark}, we know $m_{\eta}$ must satisfy 
$
m_{\eta} \delta \geq \sum\limits_{i\in \fullUserSet} \reqscalar_i \mu_i
$
, giving the following lower bound on the required number of cores: 
\begin{align}
\label{align_m_LB}
\underline{m} \equiv \Big\lceil \frac{\sum\limits_{i \in \fullUserSet} \reqscalar_i\mu_i}{\delta} \Big\rceil. 
\end{align}

\dissertationStart
If we ignore the ceilings, 
\begin{align}
\label{align_m_savings}
1 - \frac{\underline{m}}{m_{\text{RB}}} \simeq 1 - \frac{\sum\limits_{i\in \fullUserSet} \reqscalar_i \mu_i}{\sum\limits_{i\in \fullUserSet}w_i(q_i)}
\end{align}
gives us a {\em upper bound} for the possible resource savings compared with reservation-based static sharing, i.e., the percentage of cores we can save by devising the best possible non-clairvoyant resource allocation policies. Clearly, this depends on the workload distributions and the requirement vector $\reqvec$. We will see in the simulation section that the proposed approaches can achieve this upper bound in some scenarios. 
\commentEnd\fi

\subsubsection{Resource Requirements Estimate for LDF+Greedy}
~

\dissertationStart
After giving a the lower bound $\underline{m}$ for the required number of cores for all non-clairvoyant designs, we want to explore the required $m_{\text{LDF+Greedy}}$ for the LDF+Greedy policy. 
\commentEnd\fi
Ideally one would like a tight upper bound for the required resources $m_{\text{LDF+Greedy}}$ for LDF+Greedy. 
By Theorem \ref{thm_LDF_greedy_eff_ratio} we know that LDF+Greedy may expect to waste up to $\max\limits_{i \in \fullUserSet} \mu_i$ time on each core in a period because of unfinished tasks. 
\dissertationStart
Therefore, the ``effective'' time for each core in one period is at least $\delta - \max\limits_{i \in \fullUserSet} \mu_i$. 
\commentEnd\fi
Thus, to complete an ``effective'' workload $\sum\limits_{i\in \fullUserSet} \reqscalar_i \mu_i$, we propose an estimate for $m_{\text{LDF+Greedy}}$ as follows,
\begin{align}
\label{align_m_LDF_greedy_est}
m_{\text{LDF+Greedy}}^\text{est} \equiv \Big\lceil \frac{\sum\limits_{i\in \fullUserSet} \reqscalar_i \mu_i}{\delta -\max\limits_{i \in \fullUserSet} \mu_i } \Big\rceil. 
\end{align}
If $\delta \gg \max\limits_{i\in \fullUserSet}\mu_i$, this estimate is close to the lower bound $\underline{m}$.

\infocomStart
One can analytically show that indeed $m_{\text{LDF+Greedy}}^\text{est} \geq m_{\text{LDF+Greedy}}$ when $\delta$ and $n$ are large, see the extended version of this paper \cite{EXT}. 
\commentEnd\fi
We can analytically show that indeed $m_{\text{LDF+Greedy}}^\text{est} \geq m_{\text{LDF+Greedy}}$ when $\delta$ and $n$ are large, 
\tompecsonlyStart
see the extended version of this paper \cite{EXT}. 
\commentEnd\fi
\tompecsextendedStart
see the proposition as follows. 
\commentEnd\fi
We observe that the inequality holds true in the various simulation settings considered next. 

\tompecsextendedStart
\begin{proposition}
\label{proposition_m_est_for_large_system}
For a SRT-MIC  system model with homogeneous users where all users have i.i.d. NBUE task workloads with mean $\mu$ and the same QoS requirement $\reqscalar$, if the period length satisfies $1 - \frac{\mu}{\delta} > \reqscalar$, then for any NBUE workload distribution and for any $\epsilon > 0$ satisfying $1 - (1+\epsilon) \frac{\mu}{\delta} > \reqscalar$, there exists $n^\prime$, such that for all $n \geq n^\prime$, 
$$
m = \Big\lceil \frac{nq \mu}{\delta - (1+\epsilon)\mu} \Big\rceil
$$
is a sufficient number of cores to meet the QoS requirement for $n$ users. 
\end{proposition}

By letting $\epsilon$ approach $0$, the $m$ in this proposition approaches $m_{\text{LDF+Greedy}}^\text{est}$. This is due to the law of large numbers and we omit the proof.  
\commentEnd\fi

\section{Simulations}
In this section we address through simulation some of the questions that are still open: 
\begin{enumerate}
\item What are possible resource savings of adopting LDF+Greedy versus reservation-based static sharing? 
		Are they close to optimal when $\delta$ is large? 
		How do they depend on the QoS requirements $\reqvec$? 
\item Our theorems on the lower bounds on efficiency ratios imply that LDF+TS/LLREF is better than LDF+Greedy for small $\delta$ and deterministic workloads. Is it true that LDF+TS/LLREF is more efficient? 
\item For workloads with small variability, can one use LDF+TS/LLREF and get gains over LDF+Greedy?  
\end{enumerate}

Our simulation setup is as follows. We start with an initial deficit vector $\mathbf{X}(0) = (0, 0, \cdots, 0)$. In each period, we independently generate a task workload realization for each user and simulate the specified policy to evaluate if tasks complete. All simulations are run for $3000$ periods. A QoS requirement vector $\reqvec$ is feasible if for all users $i$ the fraction of task completions over the $3000$ periods exceeds $\reqscalar_i$. 

\subsection{Near-Optimality of LDF+Greedy for Large $\delta$}
\begin{figure}[htp]
  \centering
  \includegraphics[width=0.65\textwidth]{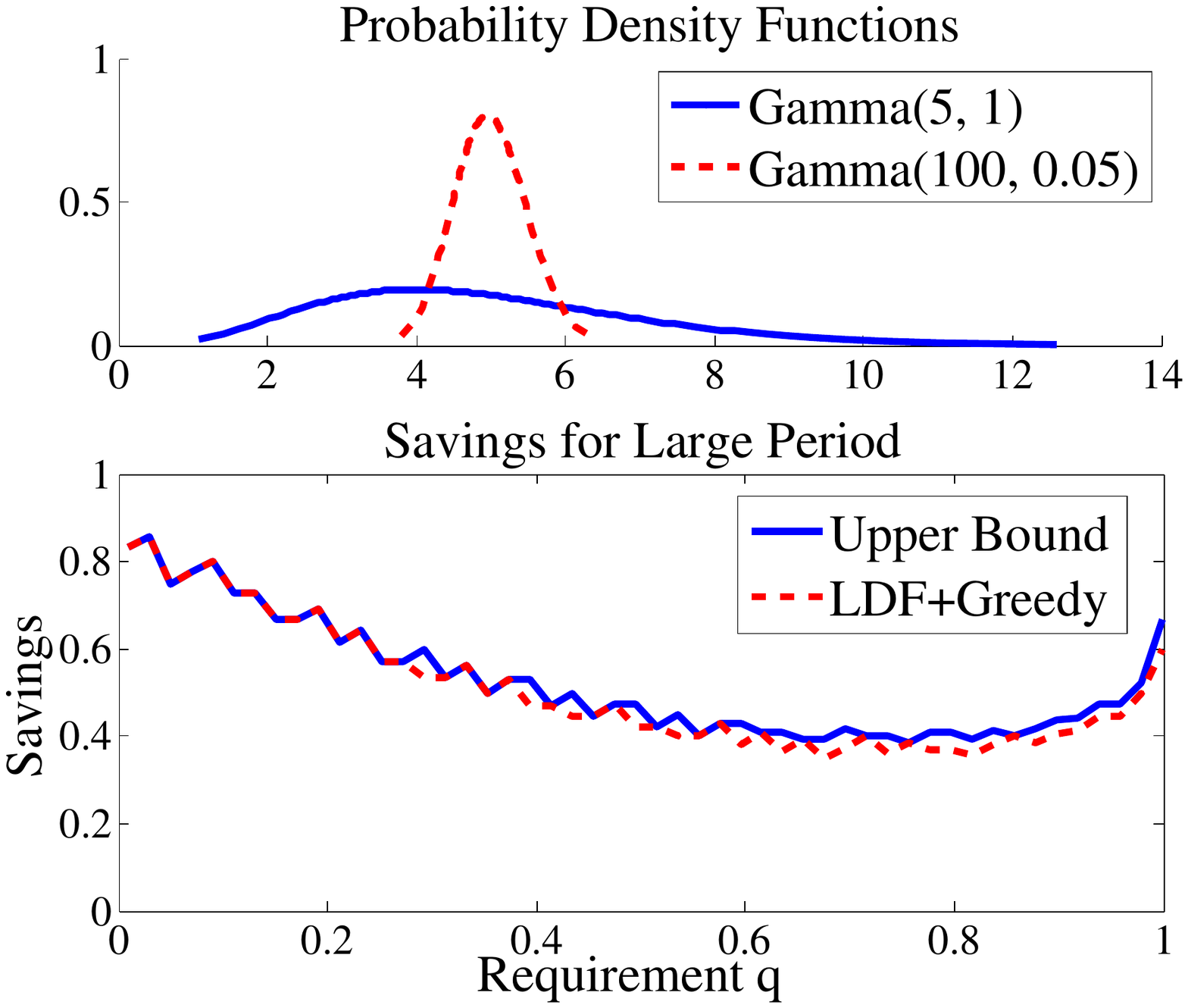}
%  \caption{Top: the probability density functions for Gamma$(5, 1)$ and Gamma$(100, 0.05)$. Bottom: the upper bound of resource savings and the actual resource savings of LDF+Greedy as $\reqscalar$ increases from $0$ to $1$.}
  \caption{Top: the probability density functions for Gamma$(5, 1)$ and Gamma$(100, 0.05)$. Bottom: the resource savings for large period.}
  \label{fig_gamma_dist_and_m_savings_large_period}
\end{figure}

\dissertationStart
LDF+Greedy is simple to implement in practice since it does not require knowledge of workload distribution or the task preemption and migration. 
\commentEnd\fi
To evaluate the resource savings of LDF+Greedy for large period length $\delta$, we consider an SRT-MIC system model with $n=200$ and $\delta=50$, serving homogeneous users that have the same QoS requirement $\reqscalar$ and generate tasks with Gamma$(5,1)$ workloads, i.e., a sum of $5$ independent exponential random variables with parameter $1$. 
The probability density function is shown in the top panel in Figure~{\ref{fig_gamma_dist_and_m_savings_large_period}}. We choose this NBUE workload distribution as a representative one. 

In the bottom panel in Figure~{\ref{fig_gamma_dist_and_m_savings_large_period}}, 
we show the simulated resource savings of LDF+Greedy versus the reservation-based static sharing, i.e., $1 - \frac{m_{\text{LDF+Greedy}}}{m_{\text{RB}}}$, and the computed upper bound on resource savings $1 - \frac{\underline{m}}{m_{\text{RB}}}$ as the QoS requirement $\reqscalar$ increases from $0$ to $1$. 
The lines are not smooth because we take ceilings when computing $\underline{m}$ and $m_{\text{RB}}$. 

It can be seen that the savings under LDF+Greedy is close to the upper bound in this setting. 
The ``U'' shape of the exhibited results depends on the workload distribution. 
Intuitively, in this homogeneous-user scenario, if we ignore the ceilings in (\ref{align_m_RB}) (\ref{align_m_LB}), the upper bound on savings becomes,
\begin{align}
\label{align_m_savings_simplified}
1 - \frac{\underline{m}}{m_{\text{RB}}} \simeq 1 - \frac{q \mu}{w(q)}, 
\end{align}
where $\mu$ is the common mean workload and $w(q)$ is the common required static allocation. 
For high $\reqscalar$, $w(q)$ is like a worst-case workload and this is an improvement from worst case to average which is as high as $60$-$70\%$ for Gamma$(5,1)$ distribution. 
For medium $\reqscalar\sim50\%$, $\reqscalar\mu$ is around $0.5\mu$ while $w(q)$ is roughly $\mu$, giving a $50\%$ resource savings. For low $\reqscalar$, $\reqscalar\mu$ is much smaller compared to $w(q)$ and the savings can be up to $80$-$90\%$. 
\dissertationStart
Intuitively, in this setting the percentage of resource savings (\ref{align_m_savings}) simplifies to
\begin{align}
\label{align_m_savings_simplified}
1 - \frac{\underline{m}}{m_{\text{RB}}} \simeq 1 - \frac{q \mu}{w(q)}, 
\end{align}
where $\mu$ is the common mean task workload and $w(q)$ represents the common required static allocation for each user. 

For SRT applications that require high $\reqscalar$, under reservation-based static sharing policies we need to allocate each user $w(q)$ which is like worst-case workload while an optimal resource allocation policy only requires $q\mu$ which is even less than $\mu$ for each user. This is an improvement from worst case to average and is as big as $60\%$-$70\%$ for Gamma$(5,1)$ distribution. 
For some applications whose QoS requirements are similar to $50\%$, e.g., the channel measurement tasks in the CRAN context, $w(q)$ is roughly $\mu$ while an optimal resource allocation policy only need around $0.5\mu$, giving a $50\%$ resource savings. 
For low $\reqscalar$, the savings can be up to $80\%$-$90\%$. 
\commentEnd\fi

%For example, under deterministic workloads these lines will be monotonically decreasing and we will discuss this in the sequel. 

% potential saving: reduce the caption. 

\subsection{LDF+Greedy vs. LDF+TS/LLREF for Deterministic Workloads and Small $\delta$} 
\dissertationStart
When period $\delta$ is comparable to mean workloads, we shall explore other resource allocation policies. 

For the SRT-MIC system model where users have the same $\reqscalar$ and generate tasks with the same deterministic workloads, by (\ref{align_m_savings_simplified}) we know $w(q)$ equals to $\mu$ and thus we get
$$
1 - \frac{\underline{m}}{m_{\text{RB}}} \simeq 1 - q. 
$$

This implies that the resource savings monotonically decrease as $q$ increases, which is different from the ``U'' shape of the lines in Figure~{\ref{fig_gamma_dist_and_m_savings_large_period}}. By Theorem \ref{thm_LDF_TS_LLREF_eff_ratio} the LDF+TS/LLREF policy provides a good option to achieve these savings. 
\commentEnd\fi

\begin{figure}[htp]
  \centering
  \includegraphics[width=0.6\textwidth]{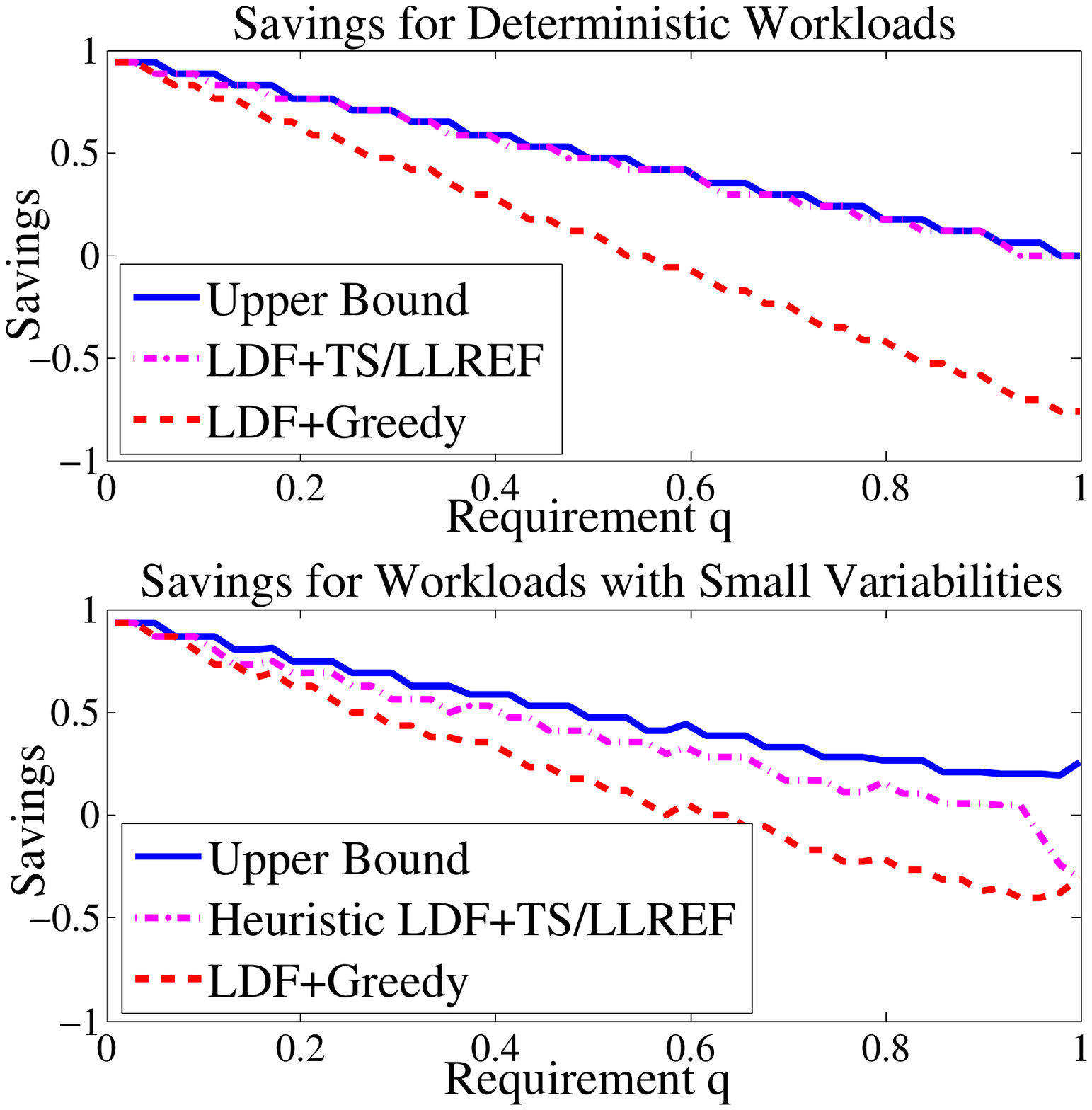}
  \caption{Top: the resource savings under deterministic workloads. Bottom: the resource savings under random workloads with small variability. }
  \label{fig_m_savings_deterministic_and_low_variability}
\end{figure}

To compare LDF+Greedy and LDF+TS/LLREF for short periods $\delta$ and deterministic workloads, we consider a system where $n = 30$ and $\delta = 9$ and where users are homogeneous and generate tasks with deterministic workloads $\mu = 5$. 
In the top panel in Figure~{\ref{fig_m_savings_deterministic_and_low_variability}}, we exhibit the upper bound of resource savings and the resource savings under LDF+Greedy and LDF+TS/LLREF as the requirement $q$ changes from $0$ to $1$. 

As can be seen, LDF+TS/LLREF can achieve the upper bound on savings while LDF+Greedy does not perform as well. 
For high $q$, the savings for LDF+Greedy is even negative implying that LDF+Greedy is worse than the reservation-based approach. 
This is because we chose $\mu$ and $\delta$ such that LDF+Greedy wastes a significant amount of time on unfinished tasks. 
Observe that the savings are monotonically decreasing in $q$, which is different from the ``U'' shape exhibited in Figure~{\ref{fig_gamma_dist_and_m_savings_large_period}}. Intuitively, this is because for deterministic workloads, by (\ref{align_m_savings_simplified}) we know $w(q)$ equals to $\mu$ and thus we get
$$
1 - \frac{\underline{m}}{m_{\text{RB}}} \simeq 1 - q. 
$$

\subsection{LDF+TS/LLREF for Workloads with Small Variability}
\label{subsubsection_heuristic_TS_LLREF_small_var}
For workloads with small variability, we envisage that the heuristic LDF+TS/LLREF described in Section \ref{subsection_LDF_TS_LLREF} is a good non-clairvoyant policy. 
\dissertationStart
For non-deterministic workloads, the LDF+TS/LLREF policy requires knowledge of workload realizations and so is not non-clairvoyant. 
However, we envisage that the heuristic estimation based LDF+TS/LLREF described in \ref{subsection_LDF_TS_LLREF} can be a good non-clairvoyant generalization, especially for random NBUE workloads with small variability. 
\commentEnd\fi
Consider a SRT-MIC system with homogeneous users where $n = 30$ and $\delta = 9$ and where the task workload distributions are Gamma$(100, 0.05)$ exhibited on the top panel in Figure~{\ref{fig_gamma_dist_and_m_savings_large_period}}. 
Note that the distribution Gamma$(100, 0.05)$ has the same mean $\mu = 5$ but a small variance. In this setting, we shall estimate the workload to be $w^\text{est} = 1.1\mu$ and use our proposed heuristic LDF+TS/LLREF in Section \ref{subsection_LDF_TS_LLREF}. 
We conduct the same analysis for resource savings and exhibit the results in the bottom panel in Figure~{\ref{fig_m_savings_deterministic_and_low_variability}}. 

As can be seen, the heuristic LDF+TS/LLREF indeed performs better than LDF+Greedy. However, the performance of the heuristic LDF+TS/LLREF degrades for high $q$. This is due to the fact that some selected tasks fail to complete since their workloads are larger than $w^\text{est}$. 
\dissertationStart
and that becomes a more critical problem as $q$ becomes bigger. 
\commentEnd\fi
One approach to solve this is to increase $w^\text{est}$ as $q$ becomes bigger. 
\dissertationStart
and we observe it can improve the resource savings for the heuristic LDF+TS/LLREF in large $q$ regime. 
\commentEnd\fi

%\begin{figure}[htp]
%  \centering
%  \includegraphics[width=0.45\textwidth]{Figures/m_savings_low_variability.pdf}
%  \caption{The resource savings under random workloads with small variabilities. }
%  \label{fig_m_savings_low_variability}
%\end{figure}

Although we only considered homogeneous users, the above observations were found to be robust for heterogeneous users. 

\dissertationStart
\subsection{Impact of Workload Heterogeneity}
In general users sharing a computing system may have heterogeneous workloads and requirements. We shall explore the impact of workload heterogeneity on our proposed designs. We consider a system similar to the one above in \ref{subsubsection_heuristic_TS_LLREF_small_var} but with heterogeneous workloads. For simplicity, we denote by HOM the system in \ref{subsubsection_heuristic_TS_LLREF_small_var} and by HET the heterogeneous system introduced below. 

\begin{figure}[htp]
  \centering
  \includegraphics[width=0.45\textwidth]{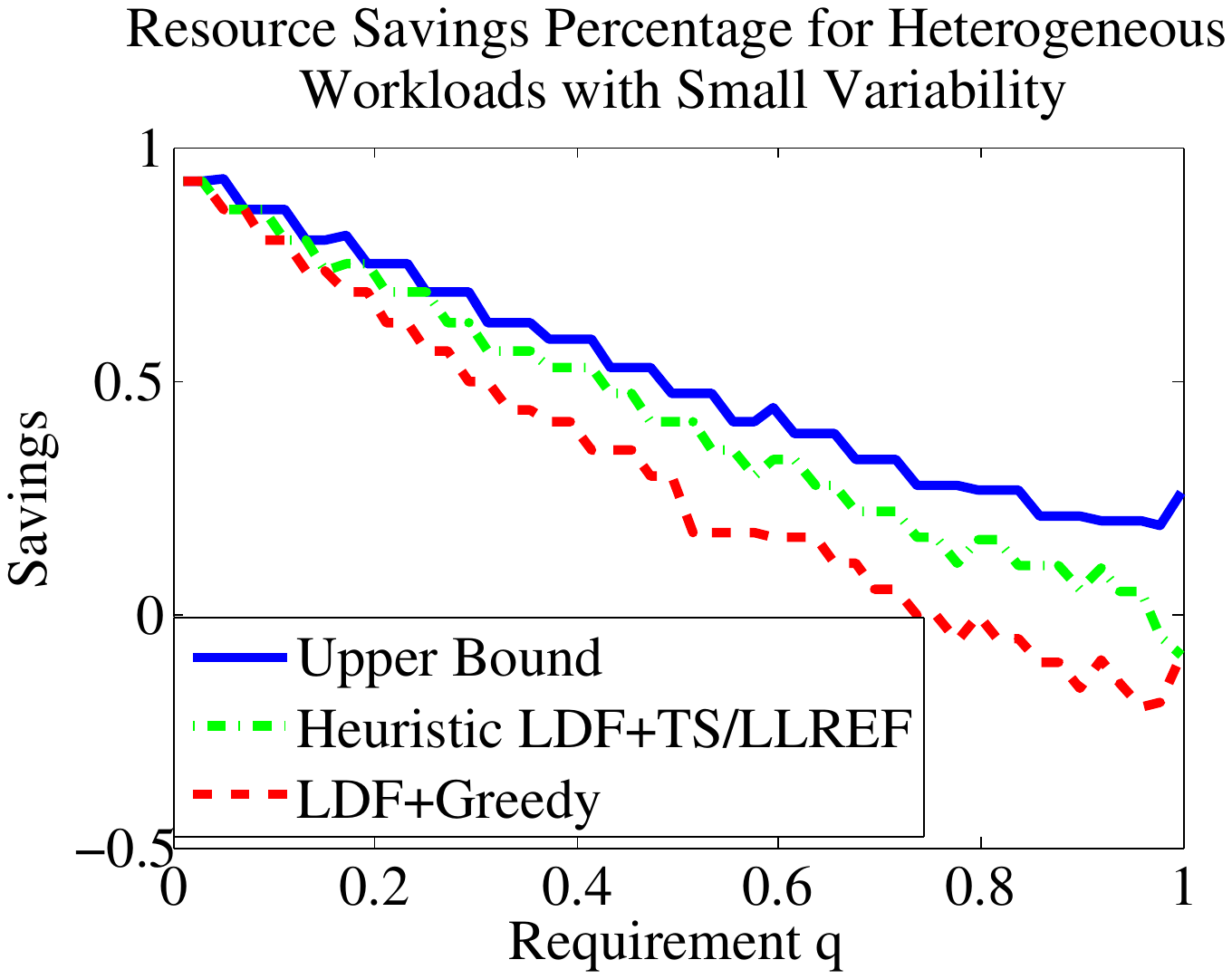}
  \caption{The resource savings under heterogeneous random workloads with small variabilities. }
  \label{fig_m_savings_heterogeneous_low_variability}
\end{figure}

In the HET system, some users in the HOM system are replaced by users with smaller task workloads. 
Specifically, the HET system has $25$ users with workload distribution being Gamma$(100, 0.05)$ and $25$ users with workload distribution being Gamma$(100, 0.01)$. Suppose the period is $\delta = 9$ and all users have the same QoS requirement $q$. By (\ref{align_m_RB}) and (\ref{align_m_LB}) it is easy to verify that with the same QoS requirement $q$ the HOM and HET systems have the same $m_{\text{RB}}$ and $\underline{m}$ and therefore, the same upper bound resource savings. 
Similarly as for the HOM system, in Figure~{\ref{fig_m_savings_heterogeneous_low_variability}} we plot the upper bound resource savings and the savings for the LDF+Greedy design and the estimation based LDF+TS/LLREF design with $W_i^\text{est} = 1.1\mu_i$ for each user $i$ in the HET system. 

We can see that the savings for the LDF+Greedy design is better in HET system implying that workload heterogeneity improves the performance of LDF+Greedy. Intuitively we know that under LDF+Greedy each core wastes time on at most $1$ unfinished task per period. By replacing big tasks with small ones, the expected wasted time decreases and thus, the performance is improved. 
The heterogeneous workloads have little impact on the performance of the heuristic LDF+TS/LLREF design. 

\commentEnd\fi

\infocomStart
\section{Generalizations to Cores with Different Processing Speeds}
\label{section_generalizations}
In this section we summarize generalizations of our results to systems consisting of cores with different processing speeds. 
Let $C = \{1, 2, \cdots, m\}$ denote the set of cores. Suppose all cores are of the same type and each core $c\in C$ has processing speed $s_c$, i.e., cores are ``uniform'', see the taxonomy in, e.g., \cite{DaB11A}. 
A task with workload $w$ processed on core $c$ has a processing time $\frac{w}{s_c}$.
Let $\overline{s} = \frac{\sum\limits_{c\in C} s_c}{m}$ be the average processing speed. 
In the model we have considered, $s_c = 1$ for all $c \in C$. 

In this setting, the outer bound $R_\text{OB}$ in Theorem \ref{thm_optimal_benchmark} becomes
\begin{align*}
R_{\text{OB}} \equiv \{ \reqvec \in \mathbb R^n_+ ~|~ \reqvec \preceq \mathbf{1}, \sum\limits_{i\in \fullUserSet} q_i\mu_i \leq  \sum\limits_{c\in C} s_c \cdot \delta\}. 
\end{align*}

For LDF+Greedy, given that cores have different processing speeds, one may want to migrate tasks to faster cores if possible. Therefore, 
depending on whether task preemption/migration is allowed, there are two types of LDF+Greedy. In Type $1$, the task scheduler greedily and preemptively schedules tasks with the highest priority on the fastest cores. In this setting, $\gamma_1$ in Theorem \ref{thm_LDF_greedy_eff_ratio} becomes
$
\gamma_1 = 1 - \frac{\max\limits_{i \in \fullUserSet} \mu_i}{\overline{s} \cdot \delta}. 
$
In Type $2$, the task scheduler greedily places tasks on available cores by priority but does not allow preemption/migration. Here $\gamma_1$ becomes
$
\gamma_1 = 1 - \frac{\max\limits_{i \in \fullUserSet} \mu_i}{\min\limits_{c \in C} s_c \cdot \delta}. 
$

We can also generalize LDF+TS/LLREF and Theorem \ref{thm_LDF_TS_LLREF_eff_ratio} with additional natural assumptions. See the extended version of this paper \cite{EXT} for more details and other generalizations.
\commentEnd\fi

\tompecsStart
\section{Possible Generalizations}
\label{section_possible_generalizations}
% to make this horizontal, use sidewaystable
\begin{table}
%\normalsize
\footnotesize
    \tbl{Results for different generalizations. \label{tab_generalizations}}{
    \centering
    \begin{tabular}{| c || c | c | c | c | c |}
    \hline
    \multirow{3}{*}{Model} & \multirow{3}{*}{Reservation-Based $\feasibilityRegion_{\text{RB}}$} & \multirow{3}{*}{Outer Bound $R_\text{OB}$} & \multicolumn{2}{|c|}{$\gamma_1$ (NBUE workloads)} & {$\gamma_2$}	\\
    \cline{4-5}
     & & & \multirow{2}{*}{preemptive} & {non-} & {(deterministic}	\\
     & & & & {preemptive} & {workloads)}	\\
    \hline\hline
    {$
	    \begin{aligned}
	    \text{SRT-}	\\
	    \text{MIC}
	    \end{aligned}
    $} 
    & {$
	    \begin{aligned}
		\{ \reqvec \in \mathbb R^n_+ ~|~ & \reqvec \preceq \mathbf{1}, 	\\
		& \sum\limits_{i \in \fullUserSet} w_i(q_i) \leq m\delta, 	\\
		& w_i(q_i) \leq \delta, \forall i \in \fullUserSet \}
	    \end{aligned}  
    $} 
    & {$
	    \begin{aligned}
		\{ \reqvec \in \mathbb R^n_+ ~|~ & \reqvec \preceq \mathbf{1}, 	\\
		& \sum\limits_{i\in \fullUserSet} q_i\mu_i \leq m\delta \}
	    \end{aligned}
    $} 
    & \multicolumn{2}{|c|}{$1 - \frac{\max\limits_{i \in \fullUserSet} \mu_i}{\delta}$} 
    & {$1 - \frac{\max\limits_{i \in \fullUserSet} \mu_i}{m\delta}$}	\\
    \hline
    {}
    & \multirow{5}{*}{$
    	\begin{aligned}
    	\{ \reqvec \in \mathbb R^n_+ ~|~ & \reqvec \preceq \mathbf{1},	\\ 
    	& B_n(\reqvec) \leq S_m \cdot \delta, 	\\
    	& B_k(\reqvec) \leq S_k \cdot \delta, 1 \leq k \leq m\}
    	\end{aligned}
    $} 
    & \multirow{5}{*}{$
    	\begin{aligned}
    	\{ \reqvec \in \mathbb R^n_+ ~|~ & \reqvec \preceq \mathbf{1}, 	\\
    	& \sum\limits_{i\in \fullUserSet} q_i\mu_i \leq  S_m \cdot \delta\}
    	\end{aligned}
    $}
    & \multirow{5}{*}{$1 - \frac{\max\limits_{i \in \fullUserSet} \mu_i}{\overline{s} \cdot \delta}$}
    & \multirow{5}{*}{$1 - \frac{\max\limits_{i \in \fullUserSet} \mu_i}{\min\limits_{c \in C} s_c \cdot \delta}$}
    & \multirow{5}{*}{$1 - \frac{\max\limits_{i \in \fullUserSet} \mu_i}{S_m \cdot \delta}$}\\
    {Different} & & & & &	\\
    {speeds} & & & & &	\\
    {$s_c$} & & & & &	\\
     & & & & &	\\
    \hline
	{}
    & \multirow{5}{*}{$
    	\begin{aligned}
    	\{ \reqvec \in \mathbb R^n_+ ~|~ & \reqvec \preceq \mathbf{1}, 	\\
    	& \sum\limits_{i \in \fullUserSet} \frac{w_i(q_i)}{\delta_i} \leq m, 	\\
    	& w_i(q_i) \leq \delta_i, \forall i \in \fullUserSet \}
    	\end{aligned}
    $} 
    & \multirow{5}{*}{$
    	\begin{aligned}
    	\{ \reqvec \in \mathbb R^n_+ ~|~ & \reqvec \preceq \mathbf{1}, 	\\
    	& \sum\limits_{i \in \fullUserSet} \frac{\reqscalar_i \mu_i}{\delta_i} \leq m \}
    	\end{aligned}
    $}
    & \multicolumn{2}{|c|}{\multirow{5}{*}{N/A}}
%    & \multirow{5}{*}{N/A} & \multirow{5}{*}{}
    & \multirow{5}{*}{$1 - \frac{\max\limits_{i \in \fullUserSet} \frac{\mu_i}{\delta_i}}{m}$}\\
    {Different} & & & \multicolumn{2}{|c|}{} &	\\
    {periods} & & & \multicolumn{2}{|c|}{} &	\\
    {$\delta_i$} & & & \multicolumn{2}{|c|}{} &	\\
     & & & \multicolumn{2}{|c|}{} &	\\
    \hline
	{} 
    & \multirow{5}{*}{$
        \begin{aligned}
    	\{ \reqvec \in \mathbb R^n_+ ~|~ & \reqvec \preceq \mathbf{1}, 	\\
    	& \sum\limits_{i \in \fullUserSet} w_i(q_i) \leq m\delta, 	\\
    	& w_i(q_i) \leq \delta, \forall i \in \fullUserSet \}
        \end{aligned}  
    $} 
    & \multirow{5}{*}{$
        \begin{aligned}
    	\{ \reqvec \in \mathbb R^n_+ ~|~ & \reqvec \preceq \mathbf{1}, 	\\
    	& \sum\limits_{i\in \fullUserSet} q_i\mu_i \leq m\delta \}
        \end{aligned}
    $} 
    & \multicolumn{2}{|c|}{\multirow{5}{*}{$1 - \frac{\max\limits_{i \in \fullUserSet} \mu_i}{\delta}$}} 
    & \multirow{5}{*}{$1 - \frac{\max\limits_{i \in \fullUserSet} \mu_i}{m\delta}$}	\\
    {Chains of} & & & \multicolumn{2}{|c|}{} &	\\
    {subtasks} & & & \multicolumn{2}{|c|}{} &	\\
    {$k(i)$} & & & \multicolumn{2}{|c|}{} &	\\
     & & & \multicolumn{2}{|c|}{} &	\\
    \hline
    \end{tabular}
    }
\end{table}

In this section we discuss the following generalizations of the SRT-MIC NBUE-workload model and associated results: 
\begin{enumerate}
\item Cores with different processing speeds. 
\item Users generating tasks at different periods. 
\item Tasks which further consist of sub-tasks that need to be processed in order. 
\end{enumerate}
We discuss these three generalizations in the following three subsections, respectively. 

For ease of reference, Table \ref{tab_generalizations} provides a summary of various generalizations\textemdash the necessary notation is introduced in the sequel.  

\subsection{Cores with Different Processing Speeds}
We first consider generalizations where the cores may have different processing speeds. 
Let $C = \{1, 2, \cdots, m\}$ denote the set of cores. Suppose all cores are of the same type and each core $c\in C$ has processing speed $s_c$, 
i.e., cores are ``uniform'', see the taxonomy in, e.g., \cite{DaB11}. 
In other words, if a task runs on a core with speed $s$ for $t$ time units, then $s\times t$ units of work are performed. 
In this context, the workload of a task refers to the required units of \emph{work} to fully complete the task. 
Therefore, a task with workload $w$ processed on core $c$ has a processing time $\frac{w}{s_c}$.
Let $\overline{s} = \frac{\sum\limits_{c\in C} s_c}{m}$ be the average processing speed. 
Clearly, in the SRT-MIC model we have previously considered, $s_c = 1$ for each $c \in C$. 

We assume $n \geq m$ since otherwise one only needs the $n$ fastest cores. 
Next we discuss generalizations of our results. 

\subsubsection{Reservation-Based Static Sharing Policies}

In reservation-based static sharing, given the computed $w_i(\reqscalar_i)$ for all users $i\in \fullUserSet$, the question is whether it is feasible to find a static allocation guaranteeing that $w_i(q_i)$ units of work can be performed for each user $i$ in each period. 

\newcommand{\sumFnName}{a}

To answer this question, we first introduce some notation. 
Given a set $Z$ of non-negative numbers and a positive integer $k$ which satisfies $1\leq k \leq |Z|$, we let $\sumFnName(Z, k)$ be the sum of the largest $k$ numbers in $Z$. 
We let $S_k = \sumFnName(\{s_c | c\in C \}, k)$. 
Given a QoS requirement vector $\reqvec$, for $1 \leq k \leq n$, we let $B_k(\reqvec) = \sumFnName(\{w_i(\reqscalar_i) | i\in \fullUserSet \}, k)$ be the sum of the $k$ largest core time reservations. 
By \cite{FGB01,FuM09}, we know that a static allocation is feasible if and only if the following conditions hold: 
\begin{align}
B_n(\reqvec) & \leq S_m \cdot \delta, \label{align_B_n_leq_S_m} \\
B_k(\reqvec) & \leq S_k \cdot \delta \text{~for~} 1 \leq k \leq m. 	\label{align_B_k_leq_S_k}
\end{align}
Intuitively, (\ref{align_B_n_leq_S_m}) implies that the sum of required reservations does not exceed the total units of work that can be performed in a period. 
And (\ref{align_B_k_leq_S_k}) implies that the $k$ largest reservation requirements can be satisfied by the $k$ fastest cores. 

Such a static allocation can be obtained according to prior work, see e.g., \cite{FGB01,FuM09}. 
Therefore, the feasibility region of reservation-based static sharing $\feasibilityRegion_\text{RB}$ is given by
\begin{align*}
\feasibilityRegion_{\text{RB}} = \{ \reqvec \in \mathbb R^n_+ ~|~ & \reqvec \preceq \mathbf{1},	B_n(\reqvec) \leq S_m \cdot \delta, 	\\
& B_k(\reqvec) \leq S_k \cdot \delta \text{~for~} 1 \leq k \leq m\}. 
\end{align*}
This is consistent with our analysis when $s_c = 1$ for all $c \in C$, see Eq (\ref{align_F_RB_MIC}). 

\subsubsection{Outer Bound $R_\text{OB}$ for the System Feasibility Region}

For a system with different core processing speeds, the outer bound $R_\text{OB}$ in Theorem \ref{thm_optimal_benchmark} needs to be modified to
\begin{align*}
R_{\text{OB}} \equiv \{ \reqvec \in \mathbb R^n_+ ~|~ \reqvec \preceq \mathbf{1}, \sum\limits_{i\in \fullUserSet} q_i\mu_i \leq  S_m \cdot \delta\},  
\end{align*}
i.e., the ``effective'' workload $\sum\limits_{i\in \fullUserSet} q_i\mu_i$ cannot exceed the maximum units of work $S_m \cdot \delta$ that can be performed in a period. 

A proof of this result requires a slight modification of that of Theorem \ref{thm_optimal_benchmark}: we replace $m\delta$ by $S_m \cdot \delta$; we redefine $U_S$ to be the total units of work performed for users in $S$ in a typical period; and we redefine $A_{i, c}$ to be the indicator random variable that user $i$'s task is unfinished and $c$ units of work are performed for user $i$'s task in a typical period. 

\subsubsection{LDF+Greedy Scheduling}

%Given an LDF-based user priority decision in a period, 
%since the cores have different processing speeds, there are two types of greedy task schedulers depending on whether task preemption/migration is allowed: 
For LDF+Greedy, if all cores have the same speed, there is no benefit of moving a running task from one core to another. 
However, if cores have different speeds, one may want to migrate tasks to faster cores if they become available. 
Therefore, depending on whether task preemption/migration is allowed, there are two types of greedy task schedulers: 
preemptive and non-preemptive greedy task scheduler. 

{\bf \em Preemptive Greedy Task Scheduler: }
In the preemptive case, the task scheduler greedily and preemptively schedules tasks with the highest priority on the fastest cores. Specifically, at all times the task scheduler guarantees that the available\footnote{A task is available if it is not completed yet. } task with the highest priority is placed on the fastest core, the available task with the second highest priority is on the second fastest core, etc. 
In this setting, similarly to Theorem \ref{thm_LDF_greedy_eff_ratio} we get the following corollary. 

\begin{corollary}
\label{corollary_gamma1_diff_speed_pre}
For the generalization of SRT-MIC model to cores with different processing speeds, the efficiency ratio of the preemptive LDF+Greedy exceeds $\gamma_1$ where
$$
\gamma_1 = 1 - \frac{\max\limits_{i \in \fullUserSet} \mu_i}{\overline{s} \cdot \delta}. 
$$
\end{corollary}

Note that in the denominator we have an average processing speed $\overline{s}$, which equals to $1$ in the SRT-MIC model we considered previously. 
Intuitively, this is because under the preemptive greedy task scheduler the unfinished tasks are always on the fastest cores. And the average processing speed of the $k$ fastest cores is at least $\overline{s}$ for $1\leq k \leq m$. 
\tompecsonlyStart
We omit the proof to save space. For a detailed proof, see the extended version of this paper \cite{EXT}. 
\commentEnd\fi
\tompecsextendedStart
Refer to Appendix \ref{appendix_pf_gamma1_diff_speed_pre} for the proof. 
\commentEnd\fi

{\bf \em Non-Preemptive Greedy Task Scheduler: }
The non-preemptive greedy task scheduler starts by putting the task with the highest priority on the fastest core, the task with the second highest priority on the second fastest core, etc. Once one of these tasks completes, it continues by processing the task with priority $m+1$ on the available core, etc. In this setting, we get the following corollary. 

\begin{corollary}
\label{corollary_gamma1_diff_speed_nonpre}
For the generalization of SRT-MIC model with different processing speeds, the efficiency ratio of the non-preemptive LDF+Greedy exceeds $\gamma_1$ where
$$
\gamma_1 = 1 - \frac{\max\limits_{i \in \fullUserSet} \mu_i}{\min\limits_{c \in C} s_c \cdot \delta}. 
$$
\end{corollary}
\tompecsonlyStart
\noindent See the extended version of this paper \cite{EXT} for the proof. 
\commentEnd\fi
\tompecsextendedStart
\noindent See Appendix \ref{appendix_pf_gamma1_diff_speed_nonpre} for the proof. 
\commentEnd\fi

Note that $\gamma_1$ under the preemptive LDF+Greedy is larger than that under the non-preemptive LDF+Greedy. This captures the benefit of task preemption/migration although these operations involve overheads in practice. 

\subsubsection{LDF+TS/LLREF Scheduling}
For deterministic workloads, we shall generalize our proposed LDF+TS/LLREF scheduling. We first introduce a further assumption. 

\begin{assumption}
\label{assumption_largest_workload_compatible}
We suppose the $n$ users' deterministic workloads are such that for all $1 \leq k \leq m$, 
$$
M_k \leq S_k \cdot \delta, 
$$
where $M_k = \sumFnName(\{\mu_i | i\in \fullUserSet \}, k)$ represents the sum of the $k$ largest workloads. 
\end{assumption}

Intuitively, this guarantees that for all $1\leq k \leq m$, the $k$ tasks with largest workloads can complete on the $k$ fastest processors in a period. 

Under Assumption \ref{assumption_largest_workload_compatible}, and by \cite{FGB01,FuM09}, we can complete all tasks in a user subset $S$ in a period by some optimal scheduling if and only if $\sum\limits_{i\in S} \mu_i \leq S_m \cdot \delta$. Such optimal scheduling algorithms include U-LLREF \cite{FuM09}, a variant of LLREF for cores with different speeds, and Proportionate Fair (Pfair) \cite{BCP96}. 

Similar to the TS/LLREF task scheduler in Definition \ref{defn_TS_LLREF}, we propose TS/U-LLREF or TS/Pfair where the task selection rule (\ref{align_j_d}) naturally becomes 
\begin{align}
\label{align_j_d_different_speed}
j(\mathbf{d}) = \max\Big\{j | \sum\limits_{i = 1}^{j} \mu_{d_i} \leq S_m \cdot \delta \Big\}, 
\end{align}
and the selected subset of users are scheduled via U-LLREF or Pfair algorithms. 

Under Assumption \ref{assumption_largest_workload_compatible}, and similarly to Theorem \ref{thm_LDF_TS_LLREF_eff_ratio}, we can show that the efficiency ratio of LDF+TS/U-LLREF or LDF+TS/Pfair exceeds $\gamma_2$ where
$$
\gamma_2 = 1 - \frac{\max\limits_{i \in \fullUserSet} \mu_i}{S_m \cdot \delta}. 
$$
The proof of this result follows that of Theorem \ref{thm_LDF_TS_LLREF_eff_ratio} by simply replacing $m\delta$ with $S_m \cdot \delta$.

\subsection{Users Generating Tasks at Different Periods}
In this subsection, we consider possible generalizations of the SRT-MIC NBUE-workload model where users generate tasks at different periods, and discuss results that cannot be generalized and/or associated difficulties. 

Specifically, suppose starting from time $0$ each user $i$ generates a task at the beginning of each period of length $\delta_i$. We assume there exists a minimum common multiple $\Delta$ of $\delta_i$ for all $i$. We shall refer to $\Delta$ as a {\em super period}. 
% Note that for those not explicitly mentioned here we assume they are the same as the SRT-MIC  system model, such as the release time and deadline of a task, the random NBUE workload assumption. 

Again, each user requires the long-term time-averaged number of tasks completed on time per period $\reqscalar_i \in [0, 1]$. To be consistent with the SRT-MIC  model, we define the feasibility in terms of the positive recurrence of a Markov chain. Given $\reqvec = (\reqscalar_1, \reqscalar_2, \cdots, \reqscalar_n)$, we keep track of the deficits of users across super periods. For each user $i\in \fullUserSet$ and super period $t+1$, we shall define deficit updates as follows,
\begin{align}
\label{align_deficit_diff_periods}
X_i(t+1) = [X_i(t) + q_i \cdot \frac{\Delta}{\delta_i} - Y_i(t+1)]^+, 
\end{align}
where $Y_i(t+1)$ is a random variable representing the number of tasks completed on time for user $i$ in super period $t+1$. Let $\mathbf{X}(t) = (X_1(t), X_2(t), \cdots, X_n(t) )$. We only consider non-clairvoyant resource allocation policies such that the process $\{\mathbf{X}(t)\}_{t\geq 1}$ is a Markov chain. A QoS requirement vector $\reqvec$ is feasible if the Markov chain $\{\mathbf{X}(t)\}_{t\geq 1}$ is positive recurrent under some non-clairvoyant resource allocation policy. 

\subsubsection{Reservation-Based Static Sharing Policies}

We first generalize the performance characterization of reservation-based static sharing policies. Similarly to the setting in {\ref{subsection_reservation_based_design}}, we can compute the required core time reservation per period $w_i(\reqscalar_i)$ for all users $i$. Now $\frac{w_i(\reqscalar_i)}{\delta_i}$ represents the required core utilization for user $i$ if we want to allocate $w_i(\reqscalar_i)$ core time to user $i$ per period. Clearly, if $\sum\limits_{i\in \fullUserSet} \frac{w_i(\reqscalar_i)}{\delta_i} > m$ we cannot meet the core time reservations $w_i(\reqscalar_i)$ for all users. Indeed, by prior work, see e.g., \cite{CRJ06,DaB11}, we can characterize the feasibility region $\feasibilityRegion_{\text{RB}}$ of reservation-based static sharing policies as follows, 
\begin{align*}
\feasibilityRegion_{\text{RB}} = \{ \reqvec \in \mathbb R^n_+ ~|~ & \reqvec \preceq \mathbf{1}, \sum\limits_{i \in \fullUserSet} \frac{w_i(q_i)}{\delta_i} \leq m, 	\\
& w_i(q_i) \leq \delta_i, \forall i \in \fullUserSet \}. 
\end{align*}
Note that this is consistent with our analysis when all users have the same period, see Eq (\ref{align_F_RB_MIC}). 

Given that $\sum\limits_{i\in \fullUserSet} \frac{w_i(\reqscalar_i)}{\delta_i} \leq m$ and $w_i(q_i) \leq \delta_i$ for all $i \in \fullUserSet$, since users have different periods, the remaining problem is how to allocate $w_i(\reqscalar_i)$ to each user $i$ in each period. 
One solution is to use the LLREF scheduling policy. 
\tompecsonlyStart
\add{We omit the details to save space. Refer to the extended version of this paper \cite{EXT} for detailed discussion. }
\commentEnd\fi
\tompecsextendedStart
Refer to Appendix \ref{appendix_LLREF_for_RB} for more details. 
\commentEnd\fi

\subsubsection{Outer Bound $R_\text{OB}$ for the System Feasibility Region}

When users generate tasks with different periods, the outer bound $R_\text{OB}$ for the system feasibility region can be generalized as follows, 
$$
R_{\text{OB}} = \{ \reqvec \in \mathbb R^n_+ ~|~ \reqvec \preceq \mathbf{1}, \sum\limits_{i \in \fullUserSet} \frac{\reqscalar_i \mu_i}{\delta_i} \leq m \}. 
$$

Intuitively, $\sum\limits_{i \in \fullUserSet} \frac{\reqscalar_i \mu_i}{\delta_i}$ represents the sum of core utilizations to fulfill QoS requirement $\reqvec$, which cannot exceed the maximum degree of parallelism $m$. The proof is similar to that of Theorem \ref{thm_optimal_benchmark}\textemdash refer to 
\tompecsonlyStart
the extended version of this paper \cite{EXT} for details. 
\commentEnd\fi
\tompecsextendedStart
Appendix \ref{appendix_pf_R_OB_diff_periods} for details. 
\commentEnd\fi

\subsubsection{LDF-Based Policies Over Super Periods}

A heuristic way to generalize our proposed LDF-based resource allocation policies to different-period scenarios is to adopt the LDF policy to pick a priority decision for each super period.  
Specifically, at the beginning of super period $t+1$, the system orders the deficit vector $\mathbf{X}(t)$ and assigns priorities from largest to smallest. 
These priorities are interpreted by the task scheduler to schedule tasks in this super period. 

{\bf \em LDF+Greedy:}
When users generate tasks with different periods, the greedy task scheduler can be preemptive or non-preemptive depending on whether preemption/migration is allowed. In the preemptive version, at all times the task scheduler processes the $m$ available tasks with the highest priority on the $m$ cores. In the non-preemptive version, the task scheduler starts with $m$ tasks with the highest priority. When a running task completes or reaches its deadline\footnote{This implies that another task from the same user is released. That new task is also considered to be a non-running task. }, the available non-running task with the highest priority is selected to be processed on the available core. 

Unfortunately, for this generalized LDF+Greedy policy we cannot get a similar performance characterization as Theorem \ref{thm_LDF_greedy_eff_ratio}. Intuitively, this is because the greedy task scheduler can potentially waste a lot of time on unfinished tasks in different-period scenarios. For example, under the preemptive greedy task scheduler, we may start processing a task right before its deadline and fail to complete it, or we may process a task only for a short time before we have to switch to process another task with higher priority leaving the original task unfinished. These scenarios degrade the performance of the LDF+Greedy policy. 

{\bf \em LDF+TS/LLREF under Deterministic Workloads: }
If the users generate tasks with different periods but with deterministic workloads, we can generalize the LDF+TS/LLREF policy and also Theorem \ref{thm_LDF_TS_LLREF_eff_ratio}. 
Naturally we assume $\mu_i \leq \delta_i$ for all $i\in \fullUserSet$. Otherwise, the tasks from user $i$ cannot complete on time. 

Under LDF+TS/LLREF, in each super period, a priority decision $\mathbf{d}$ is selected according to the LDF policy. Similarly to (\ref{align_j_d}), the system selects the user subset $J(\mathbf{d}) = \{d_1, d_2, \cdots, d_{j(\mathbf{d})}\}$ where $j(\mathbf{d})$ is computed as follows, 
\begin{align}
\label{align_j_d_generalized}
j(\mathbf{d}) = \max\Big\{j | \sum\limits_{i = 1}^{j} \frac{\mu_{d_i}}{\delta_{d_i}} \leq m \Big\}. 
\end{align}
We shall consider the case where the system adopts the LLREF policy to process and complete all tasks from $J(\mathbf{d})$ in this super period. 

To characterize the efficiency ratio, we proved the following corollary which is similar to Theorem \ref{thm_LDF_TS_LLREF_eff_ratio}. 
\begin{corollary}
\label{corollary_LDF_TS_LLREF_eff_ratio_generalized}
For the SRT-MIC  system model with different periods and deterministic workloads, the efficiency ratio of LDF+TS/LLREF that operates over super periods exceeds $\gamma_2$, where
$$
\gamma_2 = 1 - \frac{\max\limits_{i \in \fullUserSet} \frac{\mu_i}{\delta_i}}{m}. 
$$
\end{corollary}

Intuitively, under the task selection rule (\ref{align_j_d_generalized}), for the selected user subset $J(\mathbf{d})$ we know that $m - \sum\limits_{i\in J(\mathbf{d})} \frac{\mu_i}{\delta_i}$ is less than $\max\limits_{i\in \fullUserSet} \frac{\mu_i}{\delta_i}$, and therefore, the performance gap is bounded by $\frac{\max\limits_{i \in \fullUserSet} \frac{\mu_i}{\delta_i}}{m}$. The formal proof is straightforward generalization of the proof of Theorem \ref{thm_LDF_TS_LLREF_eff_ratio} and we shall omit it. 

Again, this result is consistent with our analysis when all users have the same period, see Theorem \ref{thm_LDF_TS_LLREF_eff_ratio}. 

\subsubsection{Fine-Grained LDF-Based System Designs}
A problem for the LDF-based resource allocation policies over super periods is that the task completions of users vary a lot from super period to super period. For example, a user with high priority in one super period may complete a large number of tasks in this super period and then be assigned a low priority in the next super period, completing only a small number of tasks. Such bursty completions would likely be undesirable for users especially when the super period $\Delta$ is large. 

To mitigate this problem, we could consider a fine-grained LDF policy to change the priority decisions more frequently. We divide the timeline into intervals associated with times where tasks become available for processing and deadlines. At the beginning of each interval, we compute the deficit between the QoS requirement and the actual number of completed tasks up to that time for each user $i$, sort the deficits from largest to smallest and assign priorities accordingly. 

Given the priority decision in each interval, we can adopt a greedy task scheduler. 
If task preemption/migration is allowed, naturally we start by putting the $m$ tasks with highest priority on the $m$ cores, and once one of these tasks completes, we continue by putting the task with priority $m+1$ on the available core, etc. 
If preemption/migration is not allowed, at the beginning of this interval, we continue processing the tasks running at the end of the previous interval, and once one of these tasks completes or reaches the deadline, we put the non-running task with the highest priority on the available core, etc. 
%Alternatively, if the task workloads are deterministic, we can still use the LDF+TS/LLREF approach based on the remaining workloads of the available tasks. But note that for an interval of length $\tau$, we should only select tasks with remaining workloads less than $\tau$. 

It would be of interest to characterize the performance of such resource allocation policies and to generalize LDF+TS/LLREF in future work. 

\subsection{Tasks Consisting of Sub-Tasks}
We continue our discussion of possible generalizations of our SRT-MIC NBUE-workload model to the case where each task consists of several sub-tasks that need to be processed in order and all of which need to be completed by the end of the corresponding period. We assume all sub-tasks can be processed on all cores. 

Specifically, suppose in each period each user $i\in \fullUserSet$ generates a task consisting of $k(i)$ sub-tasks, which have to be processed in order and cannot be processed in parallel. But sub-tasks of different tasks can be processed simultaneously. A task in a period is said to be completed on time if and only if all its sub-tasks complete by the end of the period. Each user $i$ requires time-averaged task completions per period $\reqscalar_i$. 
For a given user, we assume the sub-task workloads with the same sub-task index are i.i.d. across periods and the sub-task workloads with different indices are independent. 
For each user $i\in \fullUserSet$ and each sub-task index $k=1, 2, \cdots, k(i)$, we denote by $W_i^{(k)}$ the workload of the $k^\text{th}$ sub-task from user $i$ and let $\mu_i^{(k)} =  \expectation[W_i^{(k)}]$ be the mean sub-task workload. Clearly $W_i = \sum\limits_{k=1}^{k(i)} W_i^{(k)}$ and $\mu_i = \sum\limits_{k=1}^{k(i)} \mu_i^{(k)}$. 
We further assume each sub-task has an NBUE workload distribution. By \cite{ShS07b} we know user $i$'s task workload $W_i$ also has an NBUE distribution. 

This generalized task model captures tasks that are completed in phases. For example, in the CRAN context each antenna generates a task associated with each subframe. A task may further consist of sub-tasks like encoding/decoding, modulation/demodulation, FFT/IFFT.

Suppose the system can observe the sub-task completions, these observations enable a broader range of non-clairvoyant resource allocation policies, which could potentially achieve better performance, i.e., a larger system feasibility region $\feasibilityRegion$. For example, now we can consider a resource allocation policy that stops processing a task if its first sub-task takes too long. 

Clearly our original SRT-MIC  system model is a special case of this generalized model where $k(i) = 1$ for all users $i$. 
It turns out that our proposed approaches and performance characterization still hold under this generalized task model although some of the proofs need modification. Next we shall discuss this in more detail. 

\subsubsection{Reservation-Based Static Sharing Designs}

Given the sub-task workload distributions and the assumption of workload independence, we can get the workload distribution of $W_i$ and thus $w_i(q_i)$ for all users $i\in \fullUserSet$. Therefore, the discussion of reservation-based static sharing policies in Section \ref{subsection_reservation_based_design} still holds. 

\subsubsection{Outer Bound for the System Feasibility Region F}

The definition of the outer bound region $R_{\text{OB}}$ and Theorem \ref{thm_optimal_benchmark} still holds, but the proof for Theorem \ref{thm_optimal_benchmark} requires some modification. 
\tompecsonlyStart
See the extended version of this paper \cite{EXT} for the details. 
\commentEnd\fi
\tompecsextendedStart
See Appendix \ref{appendix_pf_R_OB_subtask_model} for the details. 
\commentEnd\fi

\subsubsection{LDF-Based System Designs}

We can still use our proposed LDF-based resource allocation policies, i.e., LDF+Greedy and LDF+TS/LLREF, to process tasks consisting of sub-tasks. When applying these approaches, we consider each task as a whole task and do not use the sub-task information. This is reasonable because partially completing some sub-tasks does not help to meet the QoS requirements $\reqvec$. Our performance characterization results Theorem \ref{thm_LDF_greedy_eff_ratio}, Theorem \ref{thm_LDF_TS_LLREF_eff_ratio}, etc., still hold. 

As a summary, tasks consisting of sequences of sub-tasks with independent NBUE workloads do not change the results in this paper. 

In this section we have introduced three possible generalizations in parallel. Given these results, the combinations of multiple generalizations, e.g., scenarios where the processors have different processing speeds and users generate tasks with different periods, are straightforward and we omit the discussion here. 
\commentEnd\fi

\section{Conclusion}
We have considered a computing system with multiple resources supporting soft real-time applications and established analytically and through simulation that simple resource allocation policies like LDF+Greedy are near-optimal and achieve substantial resource savings, except when the real-time constraints are tight, i.e., the period length is similar to the service time for a user's task. 
In this case, LDF+Greedy may not work well and it is worth exploring other policies. 
For workloads with small variability, we have proposed the LDF+TS/LLREF policy which indeed outperforms LDF+Greedy. 
For future work, a more detailed exploration of systems consisting of possibly different types of resources is of interest. 
%In our model we have considered multiple cores and generally distributed random workloads and in \cite{EXT} we further discuss generalizations where users may generate tasks with different periods and where a task may further consists of several sub-tasks. But our model is still restrictive in some aspects. For example, we are assuming cores are of the same type and the same speed, and the task deadlines equal to the period length. It is of great interest to extend the model to more complicated scenarios. 

\huaweiStart
\section*{Acknowledgment}
This research was supported by Huawei Technologies Co. Ltd. 
The authors would like to thank Alan Gatherer, Zheng Lu, Haishan Zhu and Mattan Erez for their comments and feedbacks on this work. 
\commentEnd\fi

\bibliography{diss_myown}{}
\bibliographystyle{ACM-Reference-Format-Journals}

\dissertationStart
\section{User Management Across SRT-MIC  Systems}
\myComments{Improve this based on prof's feedbacks. }
As we have argued, a SRT-MIC  system could be a cluster of machines or a centralized server with a large number of cores. A large-scale cloud-based infrastructure in practice generally consists of several such SRT-MIC  systems and it requires a centralized user\footnote{A user can be viewed as a stream of periodic tasks. }/stream management system to add new users to existing SRT-MIC  systems, or to move users across SRT-MIC  systems if one system is overloaded, etc. In this section, we consider a cloud-based infrastructure that consists of several SRT-MIC systems and explore how to allocate a new user to one of these systems. 

\subsection{System Model for Stream Management}
We consider a cloud-based infrastructure consisting of several SRT-MIC systems, each of which serves a set of users with NBUE workloads according to some optimized resource allocation policies, such as the LDF+Greedy policy we proposed. The users are generating tasks as we discussed in Section \ref{sec_system_model}. 
Suppose all users generate tasks with the same period $\delta$. Let $L$ be the set of all SRT-MIC systems. Each SRT-MIC system $l \in L$ has $m^{(l)}$ identical cores. For simplicity, we assume different SRT-MIC systems have identical cores, but possible with different numbers of cores. The user set for system $l \in L$ is denoted by $\fullUserSet^{(l)}$ and $n^{(l)}$ is the number of users. The QoS requirement vector for system $l \in L$ is denoted by $\reqvec^{(l)} = (\reqscalar_1^{(l)}, \reqscalar_2^{(l)}, \cdots, \reqscalar_{n^{(l)}}^{(l)})$ where $\reqscalar_i^{(l)}$ represents the required number of tasks completed on time for user $i \in N^{(l)}$. 

We can evaluate the performance of these SRT-MIC systems based on measurements of the history events. For example, for each SRT-MIC system $l\in L$ we can measure the achieved time-averaged\footnote{In practice we can measure the average over a reasonable time window. } task completion vector $\mathbf{p}^{(l)} = (p_1^{(l)}, p_2^{(l)}, \cdots, p_{n^{(l)}}^{(l)})$ where $p_i^{(l)}$ represents the achieved time-averaged number of tasks completed on time per period for user $i\in N^{(l)}$. The SRT-MIC system $l\in L$ is called feasible if $p_i^{(l)} \geq q_i^{(l)}$ for all users $i\in N^{(l)}$. Here we assume all SRT-MIC systems are feasible since otherwise we would remove users from this overloaded system. 
For each user $i \in N^{(l)}$ we can also measure the time-averaged time spent on that user per period, denoted by $t_i^{(l)}$. 

Suppose there is a new user $u$ periodically sending over tasks to this cloud-based infrastructure with QoS requirement $\reqscalar_u$ and mean workload $\mu_u$. Our objective is to allocate this new user to a SRT-MIC system aiming to meet the QoS requirement for this new user without violating the requirements of the existing users in that system. We want to ensure the QoS requirements are met while at the same time using these SRT-MIC systems efficiently. 

Based on our discussion in this paper, we have seen that it is very hard to accurately predict the impact of adding a new user to a SRT-MIC system. 
Moreover, in practice it could be challenging to get fine-grained information about workload distributions or even mean workloads since the users may not know the processing speed of the cores, the uncertainties of the SRT-MIC systems, etc. 
In this section we propose a heuristic user allocation policy that is only based on measurements of the SRT-MIC systems with the aim of motivating similar policies for more complicated cloud-based infrastructures dealing with soft real-time services. 

\subsection{Measurement-Based User Allocation Policy}
We are motivated by the inequality (\ref{align_required_workload_smaller_than_spent_time}) in the characterization of outer bound feasibility region $R_\text{OB}$. By letting $S=\{i\}$ in (\ref{align_required_workload_smaller_than_spent_time}) we know for each SRT-MIC system $l\in L$ and each user $i \in N^{(l)}$, we have
\begin{align}
\label{align_p_i_mu_i_leq_t_i}
p_i^{(l)} \mu_i \leq t_i^{(l)}, 
\end{align}
and the equality holds if user $i$ generates tasks with exponential workloads in the continuous-time scenario (or geometric workloads in the discrete-time scenario). 
Therefore, the required average time spent on user $i$ to meet QoS requirement $\reqscalar_i^{(l)}$ can be estimated by $\frac{t_i^{(l)} \reqscalar_i^{(l)}}{p_i^{(l)}}$. Next we introduce our user allocation policy. 

Our policy consists of two parts: {\em feasibility checking}, to find SRT-MIC systems that can admit this new user, and {\em scoring}, to select one out of these candidate SRT-MIC systems. Similar two-part approach is adopted in Google Borg system \cite{VPK15A} to allocate tasks, not necessarily soft real-time tasks, to machines. 

In feasibility checking, we aim to find a set of SRT-MIC systems that have enough ``space'' to admit the new user $u$. For each SRT-MIC system $l \in L$, we define the remaining space $r^{(l)}$ as follows, 
$$
r^{(l)} = m^{(l)} \delta - \sum\limits_{i\in N^{(l)}} \frac{t_i^{(l)} q_i^{(l)}}{p_i^{(l)}}. 
$$
Intuitively, $\sum\limits_{i\in N^{(l)}} \frac{t_i^{(l)} q_i^{(l)}}{p_i^{(l)}}$ represents the required time we need to spent on users in $N^{(l)}$ to meet the QoS requirement $\reqvec^{(l)}$ and $r^{(l)}$ represents the remaining time system $l$ can spend on additional users. Everything we need to compute $r^{(l)}$ is trivial to get, such as $m^{(l)}$, $\delta$, $q_i^{(l)}$, or can be measured from system $l$, i.e., $p_i^{(l)}$ and $t_i^{(l)}$. 

We mark the system $l\in L$ as a candidate system if 
$$
r^{(l)} \geq \reqscalar_u \mu_u + H, 
$$
where $H$ is a non-negative safety guard. We use $L^*$ to represent the set of candidate systems, i.e. 
$$
L^* = \{l | r^{(l)} \geq \reqscalar_u \mu_u + H \}. 
$$
We introduce this safety guard $H$ because there is a gap between $p_i^{(l)} \mu_i$ and $t_i^{(l)}$ in (\ref{align_p_i_mu_i_leq_t_i}). In the sequel we will discuss how to adapt the value of $H$ based on the ``accuracy'' of our user allocation policy. 

In scoring, we compute a score for each candidate system $l\in L^*$ according to some scoring function $f(l)$ and pick the SRT-MIC system with the largest score. We may use different scoring functions for different purposes. For example, to fill each SRT-MIC system as tight as possible, or for ``best-fit'' purpose, we can let $f(l) = -r^{(l)}$ which is equivalent to picking the system $l\in L^*$ with smallest remaining space $r^{(l)}$. That enables us to consolidate the SRT-MIC systems so that we can shut down some SRT-MIC systems to save energy or to use those resources for other purposes. But under this scoring function, we will get penalized for even small errors in the estimations of $H$, $\mu_u$, etc. 

Alternatively, to be conservative we can let $f(l) = r^{(l)}$ which is equivalent to picking the system $l \in L^*$ with the largest remaining space $r^{(l)}$. That would result in the users' workloads being balanced across SRT-MIC systems. A potential problem for using this scoring function is that the available ``space'' are separated across SRT-MIC systems and we may not be able to find a candidate system for a new user with large $\reqscalar_u \mu_u$. Some other complicated scoring functions may take into account the mixing of users with high QoS requirements and low QoS requirements. In practice we can design an appropriate scoring function based on the practical context. 

Once we allocate the new user $u$ to a SRT-MIC system $l$ according to the policy above, we check the correctness of this allocation by observing the achieved task completion vector over a reasonable time window and comparing it with the QoS requirements. If the system $l$ is no longer feasible after adding user $u$, we remove user $u$ from $l$ and re-allocate the user according to the policy above. But this time we no longer mark the system $l$ as a candidate system for user $u$. 

We want to adapt the value of safety guard $H$ based on the correctness of our user allocation decisions. Suppose $H$ starts with $\delta/2$ and we allocate user $u$ to system $l$ according to the policy above. If the user allocation decision turns out to be incorrect, that implies the value of $H$ is too small and we double the value of $H$. Otherwise, the user allocation decision is correct and we want to decrease the value of $H$ to be more aggressive. But note that $H$ only plays a role when $r^{(l)}$ is close to $\reqscalar_u\mu_u$. If $r^{(l)} \gg \reqscalar_u\mu_u$, the correctness of this decision is straightforward and does not provide much insight on whether $H$ is a good value. Therefore, in such scenario we keep $H$ unchanged. But if $r^{(l)}$ is close to $\reqscalar_u\mu_u$, by which we mean $\reqscalar_u\mu_u + 2H \geq r^{(l)} \geq \reqscalar_u\mu_u + H$, and the decision is correct, then we decrease the value of $H$ by a small constant $\epsilon$ while guaranteeing $H$ is non-negative. In this way we dynamically adapt the value of $H$ based on the correctness of user allocation decisions. 

\commentEnd\fi

\section{Appendix}

\subsection{Proof of Theorem \ref{thm_R_IB}}
\label{pf_theorem_R_IB}
We first introduce some additional notation. Given two vectors $\boldsymbol{a}=(a_1, a_2, \cdots, a_n)$ and $\mathbf{b} = (b_1, b_2, \cdots, b_n)$, we denote by $\boldsymbol{a} \circ \mathbf{b}=(a_1 b_1, a_2 b_2, \cdots, a_n b_n)$ the entrywise product. 

Given $\reqvec \in \text{int}(R_\text{IB})$, we need only show $\reqvec$ can be fulfilled by the LDF+$\mathcal{X}$ policy. 

By definition of interior there exists an $\epsilon>0$ such that ${\reqvec}^\prime = \reqvec + \epsilon\mathbf{1} \in R_{\text{IB}}$. 
By definition of $R_{\text{IB}}$, there exists a vector $\ribvec \succ \mathbf{0}$ such that for all $S \subseteq \fullUserSet$, 
\begin{align}
\label{align_q_prime_condition}
\sum\limits_{i\in S} \ribscalar_i {\reqscalar}^\prime_i \leq \min\limits_{\mathbf{d} \in D(S)} \sum\limits_{i\in S} {\ribscalar}_i p_i(\mathbf{d}). 
\end{align}

Consider the following candidate Lyapunov function: 
$$
L(\mathbf{X}(t)) = \sum\limits_{i=1}^{n} {\ribscalar}_i  X_i(t)^2. 
$$

Note that the process $\{\mathbf{X}(t)\}_{t\geq 1}$ is now driven by LDF, and let $\mathbf{Y}(t) = (Y_1(t), Y_2(t), \cdots, Y_n(t))$ be the vector of indicator variables for users' task completions under LDF. 
At period $t+1$, we have that
\begin{eqnarray}
\lefteqn{\expectation\left[L(\mathbf{X}(t+1)) - L(\mathbf{X}(t)) | \mathbf{X}(t) = \mathbf{x}\right]} \notag \\
& = & \expectation\Bigg[\sum\limits_{i=1}^{n} {\ribscalar}_i  (X_i(t+1)^2 - X_i(t)^2) | \mathbf{X}(t) = \mathbf{x}\Bigg]	\notag	\\
& \leq & \expectation\Bigg[\sum\limits_{i=1}^{n} {\ribscalar}_i  ((X_i(t) + \reqscalar_i \notag - Y_i(t+1))^2 - X_i(t)^2) | \mathbf{X}(t) = \mathbf{x}\Bigg] \notag \\
& = & \expectation\Bigg[\sum\limits_{i=1}^{n} {\ribscalar}_i  (\reqscalar_i - Y_i(t+1))^2 + 2 \langle \ribvec   \circ \mathbf{X}(t), \reqvec - \mathbf{Y}(t+1) \rangle | \mathbf{X}(t) = \mathbf{x}\Bigg] 	\notag \\
& \leq & \expectation\Bigg[\sum\limits_{i=1}^{n} {\ribscalar}_i  (\reqscalar_i^2 + Y_i(t+1)^2) + 2 \langle \ribvec   \circ \mathbf{X}(t), \reqvec - \mathbf{Y}(t+1) \rangle | \mathbf{X}(t) = \mathbf{x} \Bigg] 	\notag \\ 
& = & \expectation \Bigg[ \sum\limits_{i=1}^{n} {\ribscalar}_i  (\reqscalar_i^2 + Y_i(t+1)^2) + 2\langle \ribvec   \circ \mathbf{X}(t), {\reqvec}^\prime - \mathbf{Y}(t+1) \rangle | \mathbf{X}(t) = \mathbf{x}\Bigg] - 2\epsilon\langle \mathbf{x}, \ribvec \rangle	\label{align_LDF_drift}
\end{eqnarray}

For simplicity, let $\mathbf{d}$ denote the priority decision selected by LDF at period $t+1$. We have
\begin{eqnarray*}
\expectation\left[ \langle \ribvec   \circ \mathbf{X}(t), {\reqvec}^\prime - \mathbf{Y}(t+1) \rangle | \mathbf{X}(t) = \mathbf{x} \right] = \langle \ribvec   \circ \mathbf{x}, {\reqvec}^\prime - \mathbf{p}(\mathbf{d}) \rangle. 
\end{eqnarray*}
By reordering users according to priorities, we get
\begin{eqnarray*}
\lefteqn{\langle \ribvec   \circ \mathbf{x}, {\reqvec}^\prime - \mathbf{p}(\mathbf{d}) \rangle}	\\
& = & \sum\limits_{i=1}^{n} x_{d_i} [{\ribscalar}_{d_i} {\reqscalar}^\prime_{d_i} - {\ribscalar}_{d_i} p_{d_i}(\mathbf{d})]	\\
& = & \sum\limits_{i=1}^{n-1} [x_{d_i} - x_{d_{i+1}}] [\sum\limits_{j=1}^i {\ribscalar}_{d_j} {\reqscalar}^\prime_{d_j} - \sum\limits_{j=1}^i \ribscalar_{d_j} p_{d_j}(\mathbf{d})] + x_{d_n} [\sum\limits_{j=1}^n {\ribscalar}_{d_j} {\reqscalar}^\prime_{d_j} - \sum\limits_{j=1}^n \ribscalar_{d_j} p_{d_j}(\mathbf{d})]. 
\end{eqnarray*}

By the LDF policy we know $x_{d_i} \geq x_{d_{i+1}}$. By (\ref{align_q_prime_condition}) we have $\sum\limits_{j=1}^i {\ribscalar}_{d_j} {\reqscalar}^\prime_{d_j} \leq \sum\limits_{j=1}^i \ribscalar_{d_j} p_{d_j}(\mathbf{d})$ for $1 \leq i \leq n$. Therefore, 

$$
\expectation\left[\langle \ribvec   \circ \mathbf{X}(t), {\reqvec}^\prime - \mathbf{Y}(t+1) \rangle | \mathbf{X}(t) = \mathbf{x} \right] \leq 0. 
$$

Suppose $b$ is an upper bound for all ${\ribscalar}_i$, $\reqscalar_i$ and possible $Y_i(t+1)$, by (\ref{align_LDF_drift})
\begin{align*}
\expectation[L(\mathbf{X}(t+1)) - L(\mathbf{X}(t)) | \mathbf{X}(t) = \mathbf{x}] \leq & 2n b^3 - 2 \epsilon \langle \mathbf{x}, \ribvec \rangle \leq -1
\end{align*}
for $\mathbf{x}$ satisfying $\langle\mathbf{x}, \ribvec\rangle \geq \frac{n b^3}{\epsilon} + \frac{1}{2\epsilon}$. 

It is not hard to show\footnote{This is true because given our assumption that requirement $\reqvec$ are rational valued, the state space of process $\{\mathbf{X}(t)\}_{t\geq 1}$ is in a lattice \cite{CoS13b}.} there are finite states $\mathbf{x}$ satisfying $\langle\mathbf{x}, \ribvec\rangle < \frac{n b^3}{\epsilon} + \frac{1}{2\epsilon}$. Therefore, by Foster's Theorem $\{\mathbf{X}(t)\}_{t\geq 1}$ is positive recurrent and $\reqvec$ is fulfilled by the LDF policy. 

\subsection{Lower Bound in Theorem \ref{thm_LDF_greedy_eff_ratio} is Tight}
\label{appendix_example_showing_theorem_LDF_Greedy_eff_ratio_is_tight}
Given $\epsilon > 0$, consider a SRT-MIC system model that has $m = \lceil \frac{\frac{1}{\epsilon} + 1}{2} \rceil$ identical cores serving $2m$ users generating tasks with deterministic workload $w$ in each period of length $\delta = 2w - \frac{w}{m}$. Suppose all users have the same QoS requirement $\reqscalar$. 

In this setting, since $w \leq \delta \leq 2w$, by using LDF+Greedy one can complete $m$ tasks per period. However, by using LDF+TS/LLREF policy we can complete $\lceil \frac{m\delta}{w} \rceil = 2m-1$ tasks per period, which is a lower bound on the number of completed tasks per period under a feasibility optimal policy. 

Given that all users have the same QoS requirement, the efficiency ratio of LDF+Greedy equals to ratio of the number of tasks completed per period under LDF+Greedy to that under a feasibility optimal policy, and thus
\begin{align*}
\gamma_{\text{LDF+Greedy}} \leq \frac{m}{2m-1}.  
\end{align*}

Since $m = \lceil \frac{\frac{1}{\epsilon} + 1}{2} \rceil \geq \frac{\frac{1}{\epsilon} + 1}{2}$, we know $\epsilon \geq \frac{1}{2m-1}$. Further since $\delta = 2w - \frac{w}{m}$, we get that
$$
1 - \frac{w}{\delta} + \epsilon \geq 1 - \frac{1}{2 - \frac{1}{m}} + \frac{1}{2m-1} = \frac{m}{2m-1}. 
$$

Thus, in this setting, we have that
$$
\gamma_{\text{LDF+Greedy}} \leq 1 - \frac{w}{\delta} + \epsilon = 1 - \frac{\max\limits_{i\in \fullUserSet} \mu_i}{\delta} + \epsilon. 
$$

\subsection{Proof of Theorem \ref{thm_LDF_TS_LLREF_eff_ratio}}
\label{appendix_pf_thm_LDF_TS_LLREF_eff_ratio}
Suppose we are given a QoS requirement vector $\reqvec$. Under deterministic workloads, to fulfill $\reqvec$ the average core processing time $\sum\limits_{i\in \fullUserSet} \reqscalar_i\mu_i$ per period should not exceed $m
\delta$. Therefore, a feasible requirement vector $\reqvec$ implies 
$$
\sum\limits_{i\in \fullUserSet} \reqscalar_i \mu_i \leq m\delta, 
$$
and clearly $\reqvec \preceq \mathbf{1}$. 

The goal is to show $\gamma_2 \reqvec \in \text{cl}(F_{\text{LDF+TS/LLREF}})$. Recall that in this setting the vector $\mathbf{p}(\mathbf{d})$ represents the expected numbers of task completions per period for TS/LLREF task scheduling under priority decision $\mathbf{d}$. Given deterministic workloads and any decision $\mathbf{d}$, under LDF+TS/LLREF, $p_i(\mathbf{d})$ equals to $1$ if user $i$'s task is selected and thus completes, and equals to $0$ otherwise. By Theorem \ref{thm_R_IB} it suffices to show $\gamma_2 \reqvec \in R_{\text{IB}}$ and by letting $\ribvec = (\mu_1, \mu_2, \cdots, \mu_n)$, it suffices to show for any given user subset $S\subseteq \fullUserSet$ and priority decision $\mathbf{d} \in D(S)$, 
\begin{align}
\label{align_sum_mu_p_geq_gamma_2}
\sum\limits_{i\in S} \mu_i p_i(\mathbf{d}) \geq \gamma_2 \sum\limits_{i\in S} \mu_i \reqscalar_i. 
\end{align}

We show this in the following two cases. 

If $\sum\limits_{i\in S} \mu_i \leq m\delta$, the task selection rule (\ref{align_j_d}) will assure that all users in $S$ are selected and thus, $p_i(\mathbf{d}) = 1$ for all $i\in S$. Since $\reqvec \preceq \mathbf{1}$ and $\gamma_2 \leq 1$, we have 
$\sum\limits_{i\in S} \mu_i p_i(\mathbf{d}) = \sum\limits_{i\in S} \mu_i \geq \gamma_2 \sum\limits_{i\in S} \mu_i \reqscalar_i$. 

Otherwise, $\sum\limits_{i\in S} \mu_i > m\delta$ and then not all users in $S$ are selected. 
The task selection rule (\ref{align_j_d}) will ensure 
$$
\sum\limits_{i = 1}^{j(\mathbf{d})} \mu_{d_i} \leq m\delta < \sum\limits_{i = 1}^{j(\mathbf{d}) + 1} \mu_{d_i}
$$ and therefore, 
\begin{align*}
\sum\limits_{i\in S} \mu_i p_i(\mathbf{d}) = \sum\limits_{i = 1}^{j(\mathbf{d})} \mu_{d_i} > m\delta - \max\limits_{i\in \fullUserSet} \mu_i = m\delta (1 - \frac{\max\limits_{i\in \fullUserSet} \mu_i}{m\delta}) = \gamma_2 m\delta \geq \gamma_2 \sum\limits_{i\in \fullUserSet} \mu_i \reqscalar_i \geq \gamma_2 \sum\limits_{i\in S} \mu_i \reqscalar_i. 
\end{align*}

This proves (\ref{align_sum_mu_p_geq_gamma_2}) and therefore, 
$$
\gamma_2 \reqvec \in R_{\text{IB}} \subseteq \text{cl}(F_{\text{LDF+TS/LLREF}}). 
$$

\tompecsextendedStart
\subsection{Proof of Corollary \ref{corollary_gamma1_diff_speed_pre}}
\label{appendix_pf_gamma1_diff_speed_pre}
The proof is similar to that of Theorem \ref{thm_LDF_greedy_eff_ratio}. 
To avoid duplication here we only discuss the differences in the associated arguments. 
First, $m\delta$ should be replaced by $S_m \cdot \delta$. 
Second, instead of showing (\ref{align_wasted_smaller_than_fraction_of_total}), one needs to show
\begin{align}
\label{align_wasted_smaller_than_fraction_of_total_different_speed}
\sum\limits_{i\in S}\expectation[A_i(\mathbf{d})] \mu_i \leq \frac{\max\limits_{i \in \fullUserSet} \mu_i}{\overline{s} \cdot \delta} \expectation[T_S], 
\end{align}
for which it suffices to show that
\begin{align}
\label{align_num_unfinished_tasks_leq_different_speed}
\sum\limits_{i\in S} \expectation[A_i(\mathbf{d})] \leq \frac{\expectation[T_S]}{\overline{s} \cdot \delta}. 
\end{align}

We still define $A_S(\mathbf{d}) = \sum\limits_{i\in S} A_i(\mathbf{d})$. Under preemptive greedy task scheduling, if $A_S(\mathbf{d}) = k$ for $k = 0, 1, \cdots, m$, then there are $k$ unfinished tasks on the $k$ fastest cores, implying that the $k$ fastest cores are busy processing tasks from users in $S$ throughout the period. 
Therefore, $\sum\limits_{i\in S} W_i \geq S_k \delta$ and thus $T_S \geq S_k \cdot \delta$. Clearly by the definition of $S_k$ we know 
$$
S_1 \geq \frac{S_2}{2} \geq \cdots \frac{S_m}{m} = \overline{s}. 
$$
Therefore, $T_S \geq S_k \cdot \delta \geq k \overline{s} \delta$. 

Thus it follows that
\begin{align*}
\expectation[T_S] = \sum\limits_{k = 0}^{m} \expectation[T_S | A_S(\mathbf{d}) = k] \cdot \Pr(A_S(\mathbf{d}) = k) \geq \sum\limits_{k = 0}^{m} k\overline{s}\delta \cdot \Pr(A_S(\mathbf{d}) = k) = \overline{s}\delta \sum\limits_{i\in S} \expectation[A_i(\mathbf{d})]. 
\end{align*}

This proves (\ref{align_num_unfinished_tasks_leq_different_speed}) and concludes the proof. 
\commentEnd\fi

\tompecsextendedStart
\subsection{Proof of Corollary \ref{corollary_gamma1_diff_speed_nonpre}}
\label{appendix_pf_gamma1_diff_speed_nonpre}
The proof is similar as that of Corollary \ref{corollary_gamma1_diff_speed_pre}. But this time, instead of showing (\ref{align_wasted_smaller_than_fraction_of_total_different_speed}), we shall show
\begin{align}
\label{align_wasted_smaller_than_fraction_of_total_different_speed_nonpreemptive}
\sum\limits_{i\in S}\expectation[A_i(\mathbf{d})] \mu_i \leq \frac{\max\limits_{i \in \fullUserSet} \mu_i}{\min\limits_{c\in C} s_c \cdot \delta} \expectation[T_S], 
\end{align}
for which it suffices to show that
\begin{align}
\label{align_num_unfinished_tasks_leq_different_speed_nonpreemptive}
\sum\limits_{i\in S} \expectation[A_i(\mathbf{d})] \leq \frac{\expectation[T_S]}{\min\limits_{c \in C} s_c \cdot \delta}. 
\end{align}

This is true because, if $A_S(\mathbf{d}) = k$ for $k = 0, 1, \cdots, m$, then there are $k$ unfinished tasks, implying that there are $k$ cores busy processing tasks from users in $S$ throughout the period. 
Thus, $T_S \geq k \cdot \min\limits_{c \in C} s_c \cdot \delta$. 

Thus it follows that
\begin{align*}
\expectation[T_S] = \sum\limits_{k = 0}^{m} \expectation[T_S | A_S(\mathbf{d}) = k] \cdot \Pr(A_S(\mathbf{d}) = k) & \geq \sum\limits_{k = 0}^{m} k\cdot \min\limits_{c \in C} s_c \cdot \delta \cdot \Pr(A_S(\mathbf{d}) = k)	\\
& = \min\limits_{c \in C} s_c \cdot \delta \cdot \sum\limits_{i\in S} \expectation[A_i(\mathbf{d})], 
\end{align*}
which proves (\ref{align_num_unfinished_tasks_leq_different_speed_nonpreemptive}). 
\commentEnd\fi

\tompecsextendedStart
\subsection{Proof of $R_\text{OB}$ when users generate tasks with different periods}
\label{appendix_pf_R_OB_diff_periods}
The proof of this generalization is similar to that of Theorem \ref{thm_optimal_benchmark}. The main differences lie in the definitions of the random variables $Y_i$, $A_i$, $E_i$ and $U_S$. 
In this setting, for each user $i$ we define $Y_i$ to be the random variable that represents the number of tasks completed on time over a typical super period $\Delta$. 
For a feasible $\reqvec$, by the Ergodic Theorem, we know $q_i \cdot \frac{\Delta}{\delta_i} \leq \expectation[Y_i]$ for all $i\in \fullUserSet$. 
We further define $A_i$ to be the number of user $i$'s unfinished tasks over a typical super period and define $E_i$ to be the total residual workloads of user $i$'s unfinished tasks over a typical super period. For each subset of users $S\subseteq \fullUserSet$, we define $U_S$ to be a random variable denoting the total core time spent on users in $S$ in a typical super period. We can still get equation (\ref{align_subset_time_equation}) and by $\expectation[E_i] \leq \expectation[A_i] \mu_i$ we can get that 
$$
\sum\limits_{i\in \fullUserSet} \reqscalar_i\cdot \frac{\Delta}{\delta_i} \cdot \mu_i \leq \sum\limits_{i\in \fullUserSet}\expectation[Y_i]\mu_i \leq \expectation[U_\fullUserSet] \leq m \cdot \Delta. 
$$

Therefore, 
$$
\sum\limits_{i\in \fullUserSet} \frac{\reqscalar_i\mu_i}{\delta_i} \leq m. 
$$
\commentEnd\fi

\tompecsextendedStart
\subsection{Proof of $R_\text{OB}$ under generalized sub-task model}
\label{appendix_pf_R_OB_subtask_model}
In systems where each task consists of a sequence of sub-tasks, the definition of the outer bound region $R_{\text{OB}}$ and Theorem \ref{thm_optimal_benchmark} still holds, but the proof for Theorem \ref{thm_optimal_benchmark} requires some modification, specifically (\ref{align_E_i_total_probability}) in the proof no longer holds. 

Recall that in the proof of Theorem \ref{thm_optimal_benchmark} we want to show $\expectation[E_i] \leq \mu_i \expectation[A_i]$ for all users $i$, where $\expectation[E_i]$ is the mean residual workload of user $i$'s unfinished tasks and $\expectation[A_i]$ is the mean number of user $i$'s unfinished tasks. Our approach is to define $A_{i,c}$ to be the indicator random variable that user $i$'s task is unfinished and is processed for $c$ time units in a typical period. 
By total probability we have that 
$$
\expectation[E_i] = \sum\limits_{c = 1}^{\delta} \expectation[E_i | A_{i, c} = 1] \Pr(A_{i,c} = 1). 
$$
Under the original SRT-MIC  system model where $k(i) = 1$, by NBUE property $\expectation[E_i | A_{i, c} = 1] = \expectation[W_i - c|W_i > c] \leq \mu_i$ and that enables us to show $\expectation[E_i] \leq \mu_i \expectation[A_i]$. 

However, under this generalized task model, $\expectation[E_i|A_{i,c} = 1]$ may no longer equal to $\mu_{i,c} = \expectation[W_i - c|W_i > c]$. This is because in some resource allocation policies, the event $A_{i, c} = 1$ could give more information than $W_i > c$. For example, suppose user $i$ generates tasks with two sub-tasks, i.e., $k(i) = 2$. Consider a policy that always finishes user $i$'s sub-task $1$ and then stops. Suppose the period length $\delta$ is large enough to complete user $i$'s sub-task $1$. In this scenario we know for all $c$, $\expectation[E_i | A_{i, c} = 1] = \expectation[W_i^{(2)}]$ which may not equal to $\mu_{i,c}$. 

Next we shall show $\expectation[E_i] \leq \mu_i \expectation[A_i]$ is still true under the generalized task model for a user $i$ with $k(i)=2$. The proof can be easily extended to general $k(i)$. We define $I_i^{(1)}$ to be the indicator random variable that sub-task $1$ from user $i$ completes in a typical period. By total probability we have that
\begin{eqnarray}
\label{eqnarray_exp_E_i_given_A_i_c}
\lefteqn{\expectation[E_i | A_{i, c} = 1]}	\\
& = & \expectation[E_i | A_{i, c} = 1, I_i^{(1)} = 1] \Pr(I_i^{(1)} = 1 | A_{i,c} = 1)	 \notag \\
&  & + \expectation[E_i | A_{i, c} = 1, I_i^{(1)} = 0] \Pr(I_i^{(1)} = 0 | A_{i,c} = 1).	\notag 
\end{eqnarray}

Given that $I_i^{(1)} = 1$, the residual workload $E_i$ is only the remaining workload of sub-task $2$ and by the NBUE property of sub-task $2$, we know
$$
\expectation[E_i | A_{i, c} = 1, I_i^{(1)} = 1] \leq \mu_i^{(2)} \leq \mu_i. 
$$
Similarly, if $I_i^{(1)} = 0$, then $E_i$ is the sum of the remaining workload of sub-task $1$, and the whole workload of sub-task $2$ which is independent of the event $A_{i, c} = 1$. Therefore, by the NBUE property of sub-task $1$, we have that
\begin{align*}
\expectation[E_i | A_{i, c} = 1, I_i^{(1)} = 0] \leq \mu_i^{(1)} + \expectation[W_i^{(2)} | A_{i,c} = 1, I_i^{(1)} = 0] = \mu_i^{(1)} + \mu_i^{(2)} = \mu_i. 
\end{align*}

Now by (\ref{eqnarray_exp_E_i_given_A_i_c}) we get that
\begin{align*}
\expectation[E_i | A_{i, c} = 1] \leq \mu_i\Pr(I_i^{(1)} = 1 | A_{i,c} = 1)  + \mu_i \Pr(I_i^{(1)} = 0 | A_{i,c} = 1) \leq \mu_i. 
\end{align*}

Therefore, 
$$
\expectation[E_i | A_{i, c} = 1] \leq \sum\limits_{c = 1}^{\delta} \mu_i \Pr(A_{i,c} = 1) = \mu_i \expectation[A_i]. 
$$

The other part of the proof of Theorem \ref{thm_optimal_benchmark} remains unchanged, and thus, our discussion of $R_\text{OB}$ still holds. 
\commentEnd\fi

\tompecsextendedStart
\subsection{Achieving $\feasibilityRegion_\text{RB}$ via LLREF scheduling}
\label{appendix_LLREF_for_RB}
Given that $\sum\limits_{i\in \fullUserSet} \frac{w_i(\reqscalar_i)}{\delta_i} \leq m$ and $w_i(q_i) \leq \delta_i$ for all $i \in \fullUserSet$, since users have different periods, the challenge is how to allocate $w_i(\reqscalar_i)$ to each user $i$ in each period. 
We convert this to the following equivalent hard real-time scheduling problem. Consider a system where each user $i\in \fullUserSet$ periodically generates tasks with period $\delta_i$ and deterministic task workload $w_i(\reqscalar_i)$. The tasks are available for processing at the beginning of periods and need to be completed by the end of the corresponding periods. The objective is to schedule these tasks on $m$ identical cores to guarantee that all tasks complete on time without exception. One solution is to use the LLREF scheduling policy which always gives a feasible schedule if it is possible. In Section \ref{subsection_LDF_TS_LLREF} we have introduced LLREF policy when users have the same periods. Next we introduce how to apply LLREF to solve this hard real-time scheduling problem where users generate tasks with different periods. 

LLREF divides the timeline into intervals by task releases/deadlines. In each interval of length $\tau$, the {\em local workload} of each user $i\in \fullUserSet$ is defined as $\frac{\tau}{\delta_i} w_i(\reqscalar_i)$. Therefore, to complete all tasks on time it suffices to complete the local workloads of all users in each interval. To achieve that, in each interval we adopt the LLREF policy introduced in Definition \ref{defn_LLREF} to process local workloads for all users. This LLREF policy solves the hard real-time scheduling problem. 

By adopting this policy we can get a static time allocation such that each user $i\in \fullUserSet$ gets core time reservation $w_i(\reqscalar_i)$ in each period, which further guarantees that each user $i$ meets the QoS requirement $\reqscalar_i$. 
\commentEnd\fi

\end{document}